\documentclass[10pt,aps,prd,twocolumn,notitlepage,superscriptaddress]{revtex4-1}
\usepackage{amsthm,amsfonts,verbatim,color,times}
\usepackage{amsmath,amssymb}
\usepackage{graphicx}
\usepackage{bm}
\usepackage{epsfig,slashed}
\usepackage{amssymb}
\usepackage{amsfonts}
\usepackage{tikz}
\usepackage{xparse}
\usepackage{ifthen}
\usepackage{lipsum}
\usepackage[colorlinks=true,citecolor=blue,
linkcolor=blue,urlcolor=blue]{hyperref}
\usepackage[caption=false]{subfig}
\preprint{\mbox{}\hfill DESY 20-045 \\\mbox{}\hfill HU-EP-20/05 \\\mbox{}\hfill LTH 1231}

\begin{document}

\newcommand{\tr}{\mathop{\mathrm{tr}}}
\newcommand{\bsigma}{\boldsymbol{\sigma}}
\newcommand{\bphi}{\boldsymbol{\phi}}
\newcommand{\re}{\mathop{\mathrm{Re}}}
\newcommand{\im}{\mathop{\mathrm{Im}}}
\renewcommand{\b}[1]{{\boldsymbol{#1}}}
\newcommand{\diag}{\mathrm{diag}}
\newcommand{\sign}{\mathrm{sign}}
\newcommand{\sgn}{\mathop{\mathrm{sgn}}}

\newcommand{\halfs}{\mbox{\small{$\frac{1}{2}$}}} 
\newcommand{\Nf}{N_{\!f}}
\newcommand{\partialslash}{\partial \! \! \! /}
\newcommand{\xslash}{x \! \! \! /}
\newcommand{\yslash}{y \! \! \! /}

\newcommand{\cl}{\mathrm{cl}}
\newcommand{\mb}{\bm}
\newcommand{\ua}{\uparrow}
\newcommand{\da}{\downarrow}
\newcommand{\ra}{\rightarrow}
\newcommand{\la}{\leftarrow}
\newcommand{\mc}{\mathcal}
\newcommand{\bs}{\boldsymbol}
\newcommand{\lra}{\leftrightarrow}
\newcommand{\nn}{\nonumber}
\newcommand{\half}{{\textstyle{\frac{1}{2}}}}
\newcommand{\mf}{\mathfrak}
\newcommand{\MF}{\text{MF}}
\newcommand{\IR}{\text{IR}}
\newcommand{\UV}{\text{UV}}
\newcommand{\sech}{\mathrm{sech}}

\newcommand*{\vcenteredhbox}[1]{\begingroup
\setbox0=\hbox{#1}\parbox{\wd0}{\box0}\endgroup}
\newcommand{\picscalefactor}{0.5}


\title{Critical properties of the valence-bond-solid transition in lattice quantum electrodynamics}

\author{Nikolai Zerf}
\affiliation{Institut f\"ur Physik, Humboldt-Universit\"at zu Berlin, Newtonstra{\ss}e 15, D-12489 Berlin, Germany}

\author{Rufus Boyack}
\affiliation{Department of Physics, University of Alberta, Edmonton, Alberta T6G 2E1, Canada}
\affiliation{Theoretical Physics Institute, University of Alberta, Edmonton, Alberta T6G 2E1, Canada}

\author{Peter Marquard}
\affiliation{Deutsches Elektronen Synchrotron (DESY), Platanenallee 6, D-15738 Zeuthen, Germany}

\author{John A. Gracey}
\affiliation{Theoretical Physics Division, Department of Mathematical Sciences, University of Liverpool, P.O. Box 147, Liverpool, L69 3BX, United Kingdom}

\author{Joseph Maciejko}
\affiliation{Department of Physics, University of Alberta, Edmonton, Alberta T6G 2E1, Canada}
\affiliation{Theoretical Physics Institute, University of Alberta, Edmonton, Alberta T6G 2E1, Canada}

\begin{abstract}
Elucidating the phase diagram of lattice gauge theories with fermionic matter in 2+1 dimensions has become a problem of considerable interest in recent years, motivated by physical problems ranging from chiral symmetry breaking in high-energy physics to fractionalized phases of strongly correlated materials in condensed matter physics. For a sufficiently large number $N_f$ of flavors of four-component Dirac fermions, recent sign-problem-free quantum Monte Carlo studies of lattice quantum electrodynamics (QED$_3$) on the square lattice have found evidence for a continuous quantum phase transition between a power-law correlated conformal QED$_3$ phase and a confining valence-bond-solid phase with spontaneously broken point-group symmetries. The critical continuum theory of this transition was shown to be the $O(2)$ QED$_3$-Gross-Neveu model, equivalent to the gauged Nambu--Jona-Lasinio model, and critical exponents were computed to first order in the large-$N_f$ expansion and the $\epsilon$ expansion. We extend these studies by computing critical exponents to second order in the large-$N_f$ expansion and to four-loop order in the $\epsilon$ expansion below four spacetime dimensions. In the latter context, we also explicitly demonstrate that the discrete $\mathbb{Z}_4$ symmetry of the valence-bond-solid order parameter is dynamically enlarged to a continuous $O(2)$ symmetry at criticality for all values of $N_f$.
\end{abstract}

\maketitle

\section{Introduction}

Lattice gauge theories in 2+1 dimensions have received increasing attention in recent years. From the high-energy physics perspective, they can be viewed as a theoretical laboratory to explore ill-understood non-perturbative phenomena analogous to those of interest in four-dimensional continuum gauge theories, such as confinement~\cite{polyakov1975,polyakov1977,PolyakovBook} and chiral symmetry breaking~\cite{dagotto1989,hands1990,hands2002,hands2004,strouthos2009,karthik2016,karthik2016b}. The logic is reversed in condensed matter physics, where lattice gauge theories arise from the reparametrization of gauge-invariant, physical degrees of freedom---typically itinerant electrons or localized spins~\cite{Wen}---in terms of slave-particle or parton degrees of freedom which carry nontrivial gauge charge. Deconfined phases of lattice gauge theories arising in this context provide models of fractionalized phases of strongly correlated systems, whose unusual macroscopic properties ultimately stem from the ability of partons with fractional quantum numbers to propagate over long distances. By contrast, confinement ``glues'' the partons back together and a conventional (e.g., broken-symmetry) phase is obtained. Of particular interest is the case of fermionic partons, whose dynamics in a deconfined phase can mimic a problem of relativistic fermions interacting with dynamical gauge fields. In a variety of recently studied $\mathbb{Z}_2$ lattice gauge theories with fermionic matter~\cite{gazit2017,prosko2017,gazit2018,gazit2019,konig2019,gonzalez-cuadra2020}, some of which may be relevant to understanding the pseudogap regime of the cuprate high-temperature superconductors~\cite{sachdev2016}, a nontrivial $\mathbb{Z}_2$ flux is generated in each plaquette of the underlying square lattice, and Dirac fermions emerge at low energies, coupled to fluctuating $\mathbb{Z}_2$ gauge fields. Since discrete gauge fluctuations are necessarily gapped, these emergent Dirac fermions remain free at long distances.

By contrast, stronger effects of gauge fluctuations are expected to occur for lattice gauge theories with continuous gauge groups. Recently, sign-problem-free quantum Monte Carlo (QMC) simulations of a $U(1)$ lattice gauge theory with an even number $N_f$ of flavors of fermions on the square lattice were performed~\cite{Meng2019,Meng2019b,janssen2020}. At half filling, $\pi$ magnetic flux is spontaneously generated in each plaquette---as in the $\mathbb{Z}_2$ case---and Dirac fermions likewise emerge at low energies. The resulting model is equivalent to lattice QED$_3$ with $N_f$ flavors of four-component Dirac fermions. In contrast to the $\mathbb{Z}_2$ case, however, gapless $U(1)$ gauge fluctuations drive the ground state away from the free-Dirac fixed point. For small values of the gauge coupling, the numerical results are consistent with a gapless phase described by the deconfined, conformal QED$_3$ fixed point, which can be accessed either in the large-$N_f$ expansion~\cite{Gracey1993,gracey1994c,rantner2002,hermele2005,*hermele2007,chester2016} 
or in the $\epsilon$ expansion below four spacetime dimensions~\cite{dipietro2016,dipietro2017,dipietro2018,zerf2018}. When the gauge coupling becomes strong, a quantum phase transition from the deconfined QED$_3$ phase to a confining phase occurs, accompanied by chiral symmetry breaking and dynamical mass generation for the fermions, and is found to be continuous~\cite{Meng2019,Meng2019b,janssen2020}. For $N_f=2$, the confining phase is a N\'eel antiferromagnet. The continuum field theory of the transition, the chiral $O(3)$ QED$_3$-Gross-Neveu model, was explicitly derived in Ref.~\cite{Zerf2019} and the associated universal critical exponents were computed to four-loop order in the $\epsilon$ expansion and second order in the large-$N_f$ expansion. (A similar critical theory for an analogous transition on the kagome lattice was derived and studied at one-loop order in the $\epsilon$ expansion in Ref.~\cite{dupuis2019}.)

For $N_f=4$, $6$, and $8$, the confinement transition is found to be towards a columnar valence-bond-solid (VBS) phase with spontaneous breakdown of the $D_4$ point-group symmetry of the square lattice. The continuum field theory of the transition, the chiral $O(2)$ QED$_3$-Gross-Neveu-Yukawa (GNY) model, was explicitly derived in Ref.~\cite{boyack2019b} from the lattice gauge Hamiltonian, and shown therein to be equivalent to the gauged Nambu--Jona-Lasinio (NJL) model~\cite{Nambu1961,klevansky1989}. Critical exponents including the order parameter anomalous dimension $\eta_\phi$---which in the current context controls the asymptotic power-law decay of VBS correlations at criticality---and the correlation length exponent $\nu$ were first computed to first order in the large-$N_f$ expansion in Ref.~\cite{Gracey1993a} in arbitrary $2<d<4$ spacetime dimensions, and evaluated explicitly in 2+1 dimensions in Ref.~\cite{boyack2019b}. In the latter reference, exponents controlling the power-law decay of competing orders---charge-density-wave order (CDW), $SU(N_f)$ antiferromagnetism (AF), and quantum anomalous Hall (QAH) order---were also computed to $O(1/N_f)$ in $d=3$ dimensions. The chiral $O(2)$ QED$_3$-GNY model was also studied at one-loop order in the $\epsilon$ expansion in $d=4-\epsilon$ dimensions in Ref.~\cite{scherer2016}, where a stable fixed point was found and critical exponents computed to $O(\epsilon)$~\cite{janssen2020}.
 
In this paper we go beyond previous work on the VBS transition in lattice QED along several directions. First, we improve upon our previous large-$N_f$ study, Ref.~\cite{boyack2019b}, by computing the critical exponents $\nu$ and $\eta_\phi$ in arbitrary $2<d<4$, and performing nontrivial cross-checks with the $\epsilon$-expansion (see below). Furthermore, the order parameter anomalous dimension $\eta_\phi$ is now obtained up to $O(1/N_f^2)$. We also compute the exponent $\Delta_\text{CDW}$, which characterizes the universal power-law decay of CDW correlations at criticality, to $O(1/N_f)$. Second, we expand upon previous one-loop $\epsilon$-expansion studies. Representing the VBS order parameter by a complex scalar field $\phi$, the critical theory of the VBS transition in lattice QED is in fact not the pure $O(2)$ QED$_3$-GNY model, but contains a $\mathbb{Z}_4$ anisotropy term $\sim(\phi^4+\phi^{*4})$ similar to that appearing in critical theories of the $\mathbb{Z}_n$ clock models, and allowed by the $D_4$ point-group symmetry of the square lattice. In the large-$N_f$ limit in $d=3$, this term is irrelevant~\cite{boyack2019b}, but at finite $N_f$ in $d=4-\epsilon$ this term is relevant at tree level. (In Ref.~\cite{scherer2016}, the transition considered was the Kekul\'e VBS transition on the honeycomb lattice where instead there is a $\mathbb{Z}_3$ anisotropy $\sim(\phi^3+\phi^{*3})$.) If in fact the $\mathbb{Z}_4$ anisotropy is relevant at the fixed point of the chiral $O(2)$ QED$_3$-GNY model found in Ref.~\cite{scherer2016}, the emergent $O(2)$ symmetry would be destroyed at long distances and the transition would ultimately lie in a different universality class (or become first order). It is thus important to determine the effect of quantum corrections on the renormalization group (RG) flow of the $\mathbb{Z}_4$ anisotropy near the putative $O(2)$-symmetric QED$_3$-GNY quantum critical point (QCP), which has not been done before. Here we show by explicit calculation that the $\mathbb{Z}_4$ anisotropy is in fact an irrelevant perturbation, thus establishing the emergence of an $O(2)$ symmetry. We also improve upon existing one-loop results by computing critical exponents in the chiral $O(2)$ QED$_3$-GNY model at four-loop order in the $\epsilon$ expansion. In addition to our analytical results, we apply Pad\'e and Pad\'e-Borel resummation techniques to obtain numerical estimates of critical exponents in 2+1 dimensions for $N_f=4,6,8$, which pertain to the QMC studies mentioned earlier~\cite{Meng2019,Meng2019b,janssen2020}. Using both the large-$N_f$ and $\epsilon$ expansions we also compute the CDW exponent $\Delta_\text{CDW}$ for the chiral $O(2)$ GNY model, which characterizes the power-law decay of CDW correlations at the Kekul\'e VBS transition on the honeycomb lattice and is in principle accessible to QMC simulations such as those of Refs.~\cite{lang2013,zhou2016,li2017b,li2020}. Finally, setting $d=4-\epsilon$ we verify that our large-$N_f$ and $\epsilon$-expansion results agree order by order in the respective expansions, up to $O(\epsilon^4,1/N_f^p)$, where $p$ is one or two depending on the order at which an exponent is known. This provides strong evidence that the fermionic (GN) and bosonized (GNY) formulations of the critical theory access the same infrared fixed point, for both the gauged and ungauged models.

\section{The VBS transition in lattice QED$_3$}

We briefly review the relevant theoretical models; a more detailed discussion can be found in our previous work~\cite{boyack2019b}. The model studied numerically in Ref.~\citep{Meng2019,Meng2019b,janssen2020} is a quantum rotor model with Hamiltonian
\begin{eqnarray}
\mathcal{H} & = & \frac{1}{2}JN_{f}\sum_{\langle rr^{\prime}\rangle}\frac{1}{4}L_{rr^{\prime}}^{2}-t\sum_{i=1}^{N_{f}}\sum_{\langle rr^{\prime}\rangle}\left(c_{ri}^{\dagger}e^{i\theta_{rr^{\prime}}}c_{r^{\prime}i}+\text{h.c.}\right)\nonumber \\
 &  & +\frac{1}{2}KN_{f}\sum_{\Box}\cos\left(\boldsymbol{\Delta}\times\boldsymbol{\theta}\right).\label{eq:Hamiltonian}
\end{eqnarray}
The operators $c_{ri}^{\left(\dagger\right)}$ annihilate (create) a fermion of flavor $i=1,\dots,N_{f}$ at site $r$, where $N_{f}$ is the total number of flavors; the fermion density is fixed at $N_f/2$ fermions per site on average (half filling). The sum over $\langle rr^{\prime}\rangle$ includes only the nearest-neighbor sites $r$ and $r^{\prime}$. The angular variable $\theta_{rr^{\prime}}$ represents the coordinate operator for rotors on each bond $\langle rr^{\prime}\rangle$ of a 2D square lattice, and the eigenvalue of this operator is an element of $\left[0,2\pi\right)$.  The operator $L_{rr^{\prime}}$ is the angular momentum canonically conjugate to $\theta_{rr'}$. The term proportional to $J$ represents an electric-field contribution that governs the strength of gauge fluctuations, whereas the term proportional to $K$ is a magnetic-field contribution that favors a background flux of $\pi$ in each plaquette. The magnetic flux of each plaquette $\Box$ is defined by $\boldsymbol{\Delta}\times\boldsymbol{\theta}=\sum_{\langle rr^{\prime}\rangle\in\Box}\hat{\theta}_{rr^{\prime}}$, where the summation over $\hat{\theta}_{rr^{\prime}}$ is taken round the elementary plaquette. 

In the absence of gauge fluctuations, i.e., when $J=0$, the angular variable $\theta_{rr^{\prime}}$ is a classical variable with no imaginary-time dynamics. The background gauge flux is $\pi$, as a consequence of Lieb's theorem~\cite{lieb1994} and the positive sign of the magnetic coupling $K>0$, which produces two two-component Dirac fermions (or alternatively, one four-component Dirac fermion $\Psi_i$) per flavor $i$ in the single-particle fermion spectrum. Turning on a nonzero value of $J$ produces gauge fluctuations $A_\mu$ which minimally couple to the Dirac fermions $\Psi_i$. In the QMC simulations, a critical value $J=J_c(N_f)$ dependent on $N_f$ is found such that, for $J<J_c(N_f)$, the ground state is described by the conformal QED$_3$ fixed point, and, for $J>J_c(N_f)$, the ground state is a confined VBS phase (we focus on $N_f=4,6,8$). The continuum theory of the transition was derived from the lattice Hamiltonian in Ref.~\cite{boyack2019b}, and is of the form
\begin{eqnarray}
\mathcal{L} & = & \sum_{i=1}^{N_{f}}\overline{\Psi}_{i}\left[\slashed{D}+ig(\phi_1\Gamma_3+\phi_2\Gamma_5)\right]\Psi_{i}\nonumber\\
& & +\frac{1}{4}F_{\mu\nu}^{2}+\frac{1}{2\xi}\left(\partial_{\mu}A_{\mu}\right)^{2}+\mathcal{L}_{\phi}.\label{eq:XYQEDGN_Model}
\end{eqnarray}
We define the following $4\times 4$ Euclidean gamma matrices,
\begin{equation}
\Gamma_{\mu}=\left(\begin{array}{cc}
\widetilde{\gamma}_{\mu} & 0\\
0 & -\widetilde{\gamma}_{\mu}
\end{array}\right),\ \mu=0,1,2,
\end{equation}
and
\begin{equation}
\Gamma_{3}=\left(\begin{array}{cc}
0 & -i\\
i & 0
\end{array}\right),\ \Gamma_{5}=\Gamma_{0}\Gamma_{1}\Gamma_{2}\Gamma_{3}=\left(\begin{array}{cc}
0 & 1\\
1 & 0
\end{array}\right),
\end{equation}
where $\widetilde{\gamma}_{\mu}=\left(\sigma_{3},\sigma_{2},-\sigma_{1}\right)$
are $2\times2$ Euclidean Dirac matrices, and $\sigma_{1,2,3}$ are the usual Pauli matrices. We define the Dirac conjugate field $\overline{\Psi}_i=\Psi_i^{\dagger}\Gamma_{0}$, the gauge-covariant derivative $\slashed{D}=\Gamma_\mu(\partial_\mu+ieA_\mu)$, and the field strength tensor $F_{\mu\nu}=\partial_\mu A_\nu-\partial_\nu A_\mu$; $\xi$ is a gauge-fixing parameter.

The scalar fields $\phi_1$ and $\phi_2$ represent the $x$ and $y$ components of the fluctuating VBS order parameter, respectively. Combining them into a complex scalar field $\phi=\phi_1+i\phi_2$, their dynamics is governed by the Lagrangian
\begin{equation}
\mathcal{L}_{\phi}=\frac{1}{2}|\partial_{\mu}\phi|^{2}+\frac{1}{2}m^{2}|\phi|^{2}+\lambda^{2}|\phi|^{4}.
\end{equation}

The model thus defined is known as the chiral $O(2)$ QED$_3$-GNY model~\cite{Meng2019}, since it possesses a global $SO(2)$ symmetry under $\phi\rightarrow e^{i\theta}\phi$, $\Psi_i\rightarrow e^{-iW\theta/2}\Psi_i$, where $W=-i\Gamma_3\Gamma_5$. Apart from an anisotropy term discussed below, it is identical to the field theory considered in Ref.~\cite{scherer2016} for the Kekul\'e VBS transition on the honeycomb lattice in the presence of a dynamical gauge field. The scalar-field content and form of the Yukawa coupling in Eq.~(\ref{eq:XYQEDGN_Model}) make it differ from both the chiral Ising QED$_3$-GNY model~\cite{gracey1992,gracey1993c,janssen2017,ihrig2018,
zerf2018,gracey2018,gracey2018b,Boyack2019}, which involves a single real scalar field coupled to a fermion mass bilinear $\sum_i\overline{\Psi}_i\Psi_i$, and the chiral $O(3)$ or Heisenberg QED$_3$-GNY model~\cite{Zerf2019,dupuis2019}, which possesses a triplet of scalar fields coupled to a fermion spin bilinear $\sum_i\overline{\Psi}_i\boldsymbol{\sigma}\Psi_i$ where $\boldsymbol{\sigma}$ denotes a vector of spin Pauli matrices. It also differs from a ``Higgs-QED$_3$-GNY'' model studied in Ref.~\cite{boyack2018}, also with a single complex scalar field, but where the $U(1)$ symmetry is the gauge symmetry and the scalar field is minimally coupled to the gauge field with gauge charge twice that of the fermion field. The conformal QED$_3$ phase of the lattice gauge theory (\ref{eq:Hamiltonian}) corresponds to the unbroken phase of the chiral $O(2)$ QED$_3$-GNY model (\ref{eq:XYQEDGN_Model}), where $\langle\phi\rangle=0$, while the VBS phase is the broken phase with $\langle\phi\rangle\neq 0$. However, the symmetry broken in the VBS phase is really a discrete $C_4\cong\mathbb{Z}_4$ rotation symmetry with $\theta=\pi k/2$, $k=0,\ldots,3$. In the continuum theory, this discrete symmetry allows for a coupling of the form
\begin{align}\label{Lanis}
\mathcal{L}_\text{anis.}=b(\phi^4+\phi^{*4}),
\end{align}
in the long-wavelength critical theory. Many other couplings are allowed by symmetries, but Eq.~(\ref{Lanis}) is the only $\mathbb{Z}_4$ anisotropy term that is relevant or marginal near four dimensions. This implies that only the gauge coupling $e^2$, the Yukawa coupling $g^2$, the $|\phi|^4$ coupling $\lambda^2$, the anisotropy $b$, and the scalar field mass squared $m^2$ need to be kept in the RG analysis to follow.

\section{$\epsilon$ Expansion}
\label{sec:epsilon}

We first set the anisotropy coupling $b$ in Eq.~(\ref{Lanis}) to zero, and in Sec.~\ref{sec:Z4anis} we return to its effect on the critical properties. In order to perform a perturbative RG analysis of Eq.~\eqref{eq:XYQEDGN_Model}, its extension to arbitrary $d$ dimensions is required.  To facilitate the dimensional continuation of the chiral $O(2)$ QED$_3$-GNY model, which was derived from a lattice gauge theory in fixed 2+1 dimensions, it is convenient to first rewrite it in terms of the equivalent gauged NJL model~\cite{boyack2019b}. We introduce a new set of gamma
matrices defined by 
\begin{align}
\gamma_{\mu}&=i\Gamma_{\mu}\Gamma_{3},\,\mu=0,1,2,\\
\gamma_3&=\Gamma_3,\\
 \gamma_{5}&=-i\Gamma_{3}\Gamma_{5}.
\end{align}
In addition, define $\psi_{i}=\Psi_{i}$ and $\overline{\psi}_{i}=\Psi_{i}^{\dagger}\gamma_{0}$.
Performing these transformations, we obtain the gauged NJL Lagrangian~\cite{Nambu1961,klevansky1989},
\begin{eqnarray}
\mathcal{L} & = & \sum_{i=1}^{N_{f}}\overline{\psi}_{i}
\left[\slashed{D}+g\left(\phi_{1}+i\phi_{2}\gamma_{5}\right)\right]\psi_i
+\frac{1}{4}F_{\mu\nu}^{2}+\frac{1}{2\xi}\left(\partial_{\mu}A_{\mu}\right)^{2}\nonumber \\
&  & +\frac{1}{2}|\partial_{\mu}\phi|^{2}+\frac{1}{2}m^{2}|\phi|^{2}+\lambda^{2}|\phi|^{4},\label{eq:Lagrangian_NJL}
\end{eqnarray}
where the gauge-covariant derivative is now defined as $\slashed{D}=\gamma_\mu(\partial_\mu+ieA_\mu)$. The $SO(2)$ symmetry of the $O(2)$ QED-GNY model is realized as a global chiral
$U(1)$ symmetry of the gauged NJL model: $\psi_{i}\rightarrow e^{-i\gamma_{5}\theta/2}\psi_{i}$,
with a concomitant $U(1)$ rotation of the scalar field $\phi\rightarrow e^{i\theta}\phi$ as before.
In the absence of gauge coupling ($e^2=0$), Eq.~(\ref{eq:Lagrangian_NJL})
reduces to the ordinary (ungauged) NJL model, or equivalently the chiral $O(2)$ or XY GNY model, which was previously studied in the $\epsilon$ expansion up to four-loop order~\citep{rosenstein1993,Zerf2017}.

To study the critical properties (i.e., the $m^{2}=0$ limit) of the model~\eqref{eq:Lagrangian_NJL} in $d=4-\epsilon$ space-time dimensions, we use field-theoretic RG and the modified minimal subtraction ($\overline{\text{MS}}$) prescription.
In terms of bare fields $\psi_{i}^{0},\phi_{0},A_{\mu}^{0}$ and bare coupling constants $e_{0},\xi_{0},m_{0},\lambda_{0},g_{0}$, with $\slashed{D}^{0}=\gamma_{\mu}\left(\partial_{\mu}+ie_{0}A_{\mu}^{0}\right)$ and $\phi_{0}=\phi^{0}_{1}+i\phi^{0}_{2}$, the bare Lagrangian is written as 
\begin{eqnarray}
\mathcal{L}_{0} & = & \sum_{i=1}^{N_{f}}\overline{\psi}_{i}^{0}
\left[\slashed{D}^{0}+g_{0}\left(\phi^{0}_{1}+i\phi^{0}_{2}\gamma_{5}\right)\right]\psi_{i}^{0}\nonumber\\
&  & +\frac{1}{2}|\partial_{\mu}\phi_{0}|^{2}+\frac{1}{2}m_{0}^{2}|\phi_{0}|^{2}+\lambda_{0}^{2}|\phi_{0}|^4\nonumber\\
& &+\frac{1}{4}\left(F_{\mu\nu}^{0}\right)^{2}+\frac{1}{2\xi_{0}}\left(\partial_{\mu}A_{\mu}^{0}\right)^{2}.
\end{eqnarray}
The renormalized Lagrangian, written in terms of renormalized fields
$\psi_{i}=Z_{\psi}^{-1/2}\psi_{i}^{0},A_{\mu}=Z_{A}^{-1/2}A_{\mu}^{0}$,
$\phi=Z_{\phi}^{-1/2}\phi_{0}$, and covariant derivative $\slashed{D}=\gamma_{\mu}\left(\partial_{\mu}+ie\mu^{\epsilon/2}A_{\mu}\right)$, is 
\begin{eqnarray}
\mathcal{L}_{R} & = & \sum_{i=1}^{N_{f}}\overline{\psi}_{i}
\left[Z_{\psi}\slashed{D}+Z_{g}g\mu^{\epsilon/2}\left(\phi_{1}+i\phi_{2}\gamma_{5}\right)\right]\psi_{i}
\nonumber\\
&  & +\frac{1}{2}Z_{\phi}|\partial_{\mu}\phi|^{2}+\frac{1}{2}Z_{\phi^{2}}\mu^{2}m^{2}|\phi|^{2}+Z_{\lambda^{2}}\lambda^{2}\mu^{\epsilon}|\phi|^{4}\nonumber\\
& &+\frac{1}{4}Z_{A}F_{\mu\nu}^{2}+\frac{1}{2\xi}\left(\partial_{\mu}A_{\mu}\right)^{2}.
\end{eqnarray}
The dimensionless renormalized coupling constants are
\begin{eqnarray}
e^{2} & = & e_{0}^{2}\mu^{-\epsilon}Z_{A},\label{eq:e2}\\
g^{2} & = & g_{0}^{2}\mu^{-\epsilon}Z_{\Psi}^{2}Z_{\phi}Z_{g}^{-2},\label{eq:g2}\\
\lambda^{2} & = & \lambda_{0}^{2}\mu^{-\epsilon}Z_{\phi}^{2}Z_{\lambda^{2}}^{-1},\label{eq:l2}\\
m^{2} & = & m^{2}_{0}\mu^{-2}Z_{\phi}Z^{-1}_{\phi^{2}},\\
\xi & = & \xi_{0}Z^{-1}_{A},
\end{eqnarray}
where $\mu$ is an arbitrary renormalization scale. The renormalization
constants $Z_{X}$, $X=\psi,A,\phi,\phi^{2},\lambda^{2},g$ are calculated
up to four-loop order using an automated setup, the technical details
of which can be found in previous publications~\cite{Zerf2017,zerf2018,Zerf2019}. We perform computations in an arbitrary $\xi$-gauge, which allows us to explicitly verify that all anomalous dimensions defined below [Eq.~(\ref{gammaX})] are properly gauge-invariant (except that for the fermion field $\psi$, which is not a gauge-invariant operator).
The number of diagrams that arise during the perturbative calculation of the renormalization
constants is significantly larger than for the chiral Ising and chiral $O(3)$ QED-GNY theories~\cite{zerf2018,Zerf2019}, as in the pure GNY theories~\cite{zerf2018b}. The difference arises from the fact that in the Ising and $O(3)$ theories, the Yukawa vertex insertions are either proportional to the identity or to a spin Pauli matrix, both of which commute with the gamma matrices $\gamma_\mu$ that appear in numerator traces (coming from fermion propagators and QED vertex insertions). Traces over spin Pauli matrices and gamma matrices factorize and can be evaluated independently. By contrast, no such factorization occurs in the ungauged or gauged NJL models, since the Yukawa vertex contains a $\gamma_5$ matrix which anticommutes with $\gamma_\mu$.

\subsection{Beta functions}

The beta functions for the coupling constants $\alpha=e,g,\lambda$
are defined by 
\begin{equation}
\beta_{\alpha^{2}}=\mu\frac{d\alpha^{2}}{d\mu}.
\end{equation}
Rescaled couplings, where $\alpha^{2}/\left(4\pi\right)^{2}\rightarrow\alpha^{2}$,
are used throughout the paper. Using the definitions in Eqs.~\eqref{eq:e2}-\eqref{eq:l2},
along with the fact that the bare coupling constants are independent
of $\mu$, the beta functions become 
\begin{eqnarray}
\beta_{e^{2}} & = & \left(-\epsilon+\gamma_{A}\right)e^{2},\label{eq:beta_e2}\\
\beta_{g^{2}} & = & \left(-\epsilon+2\gamma_{\psi}+\gamma_{\phi}-2\gamma_{g}\right)g^{2},\label{eq:beta_g2}\\
\beta_{\lambda^{2}} & = & \left(-\epsilon+2\gamma_{\phi}-\gamma_{\lambda^{2}}\right)\lambda^{2},\label{eq:beta_lambda2}
\end{eqnarray}
where the anomalous dimensions associated with renormalization constants
$Z_{X}$, $X=\psi,A,\phi,\phi^{2},\lambda^{2},g$ are defined by
\begin{equation}\label{gammaX}
\gamma_{X}=\mu\frac{d\ln Z_{X}}{d\mu}.
\end{equation}
Whilst Eqs.~\eqref{eq:beta_e2}-\eqref{eq:beta_lambda2} have the same
form as in the chiral Ising~\cite{zerf2018} and chiral $O(3)$~\cite{Zerf2019} QED-GNY models,
the explicit form of the renormalization constants, and
thus the form of the anomalous dimensions, are different for the three
theories. As in the aforementioned articles, the four-loop beta functions can be expressed as a 
sum over contributions at a fixed loop order:
\begin{equation}
\beta_{\alpha^{2}}=-\epsilon\alpha^{2}+\beta_{\alpha^{2}}^{\left(1\text{L}\right)}+\beta_{\alpha^{2}}^{\left(2\text{L}\right)}+\beta_{\alpha^{2}}^{\left(3\text{L}\right)}+\beta_{\alpha^{2}}^{\left(4\text{L}\right)}.
\end{equation}
Here we present our results up to and including three-loop order;
the four-loop contributions are lengthy and are deferred to the Supplemental Material~\cite{SuppMat}. The beta functions $\beta_{e^{2}}$
for the gauge coupling are given by
\begin{eqnarray}
\beta_{e^{2}}^{\left(1\text{L}\right)} & = & \frac{8}{3}N_{f}e^{4},\label{eq:betae21}\\
\beta_{e^{2}}^{\left(2\text{L}\right)} & = & 8N_{f}e^{6}-4N_{f}e^{4}g^{2},\label{eq:betae22}\\
\beta_{e^{2}}^{\left(3\text{L}\right)} & = & -6N_{f}e^{6}g^{2}+N_{f}\left(7N_{f}+6\right)e^{4}g^{4}\nonumber \\
 &  & -\frac{4}{9}N_{f}\left(22N_{f}+9\right)e^{8}.\label{eq:betae23}
\end{eqnarray}
Similarly, the beta functions $\beta_{g^{2}}$ for the Yukawa coupling
are given by
\begin{eqnarray}
\beta_{g^{2}}^{\left(1\text{L}\right)} & = & 2g^{4}\left(N_{f}+1\right)-12e^{2}g^{2},\label{eq:betag21}\\
\beta_{g^{2}}^{\left(2\text{L}\right)} & = & -\frac{1}{2}\left(12N_{f}-7\right)g^{6}+2\left(5N_{f}+8\right)e^{2}g^{4}\nonumber \\
 &  & +\frac{2}{3}\left(20N_{f}-9\right)e^{4}g^{2}-64g^{4}\lambda^{2}+128g^{2}\lambda^{4},\label{eq:betag22}\\
\beta_{g^{2}}^{\left(3\text{L}\right)} & = & \frac{1}{8}g^{8}\left[N_{f}\left(52N_{f}+48\zeta_{3}+15\right)+48\zeta_{3}-227\right]\nonumber \\
 &  &  +\frac{2}{27}e^{6}g^{2}\left[N_{f}\left(280N_{f}-108(24\zeta_{3}-23)\right)-3483\right]\nonumber \\
 &  & -\frac{1}{2}e^{4}g^{4}\left[N_{f}\left(32N_{f}+432\zeta_{3}-33\right)+144\zeta_{3}-157\right]\nonumber \\
 &  & +64e^{2}g^{4}\lambda^{2}-e^{2}g^{6}(27N_{f}+70) -2560g^{2}\lambda^{6} \nonumber \\
 &  & +48g^{6}\lambda^{2}(5N_{f}+6)-480g^{4}\lambda^{4}(N_{f}-3).\label{eq:betag23}
\end{eqnarray}
Finally, the beta functions $\beta_{\lambda^{2}}$ for the four-scalar
coupling are given by
\begin{eqnarray}
\beta_{\lambda^{2}}^{\left(1\text{L}\right)} & = & 80\lambda^{4}+4N_{f}g^{2}\lambda^{2}-\frac{1}{2}N_{f}g^{4},\label{eq:betal21}\\
\beta_{\lambda^{2}}^{\left(2\text{L}\right)} & = & -3840\lambda^{6}-2N_{f}e^{2}g^{4}+20N_{f}e^{2}g^{2}\lambda^{2}\nonumber\\
 &&+2N_{f}g^{4}\lambda^{2}-160N_{f}g^{2}\lambda^{4}+2N_{f}g^{6},\label{eq:betal22}\\
\beta_{\lambda^{2}}^{\left(3\text{L}\right)}&=& 512\lambda^{8}\left(384\zeta_{3}+617\right)\nonumber\\
&& -\frac{1}{16}N_{f}g^{8}\left(154N_{f}+96\zeta_{3}-53\right)\nonumber\\
&&+\frac{1}{4}N_{f}e^{4}g^{4}\left(116N_{f}-96\zeta_{3}+131\right)\nonumber\\
&&-N_{f}e^{4}g^{2}\lambda^{2}\left(32N_{f}-144\zeta_{3}+119\right)\nonumber\\
&&+\frac{1}{2}N_{f}e^{2}g^{6}\left(48\zeta_{3}-7\right)+N_{f}e^{2}g^{4}\lambda^{2}\left(373-528\zeta_{3}\right)\nonumber\\
&&+120N_{f}e^{2}g^{2}\lambda^{4}\left(16\zeta_{3}-17\right)+7232N_{f}g^{2}\lambda^{6}\nonumber\\
&&+\frac{1}{4}N_{f}g^{6}\lambda^{2}\left(512N_{f}-336\zeta_{3}-1339\right)\nonumber\\
&&-4N_{f}g^{4}\lambda^{4}\left(60N_{f}-408\zeta_{3}-509\right).\label{eq:betal23}
\end{eqnarray}
In these expressions $\zeta_{3}\equiv\zeta(3)\approx1.202$ is Ap$\acute{\textrm{e}}$ry's constant. 

The results for the beta functions can be compared against existing results in the literature for specific cases. In  the limit $g^{2}=\lambda^{2}=0$ the model reduces to pure QED with $N_{f}$ flavors of four-component Dirac fermions.  The results in Eqs.~\eqref{eq:betae21}-\eqref{eq:betae23}, together with the four-loop result~\cite{SuppMat}, agree with the four-loop QED beta function~\cite{gorishny1987}.  In the limit $e^{2}=g^{2}=0$ the theory reduces to the $O(2)$ vector model; the beta functions in Eqs.~\eqref{eq:betal21}-\eqref{eq:betal23} and Ref.~\cite{SuppMat} agree in that limit with the four-loop beta function for that model~\cite{vladimirov1979}. Finally, in the limit $e^{2}=0$ the model reduces to the ungauged NJL or chiral $O(2)$/XY GNY model,  and the results in Eqs.~\eqref{eq:betag21}-\eqref{eq:betag23} and Ref.~\cite{SuppMat} agree with the four-loop result in Ref.~\cite{Zerf2017}.

\subsection{Quantum critical point}
\label{sec:QCP}

We now utilize the beta functions obtained in the previous
section to investigate the existence of a QCP for the VBS transition, which should correspond to a stable RG fixed point within the critical ($m^2=0$) hypersurface. At one-loop order, our beta functions for the gauge coupling $e^2$, the Yukawa coupling $g^2$, and the four-scalar coupling $\lambda^2$ agree with those derived in Ref.~\cite{scherer2016}. Denoting the critical couplings by $\left(e_{*}^{2},g_{*}^{2},\lambda_{*}^{2}\right)$, we thus find the same eight fixed points as them: the Gaussian fixed point $\left(0,0,0\right)$, the conformal QED fixed point $\left(\frac{3\epsilon}{8N_{f}},0,0\right)$, the $O(2)$ Wilson-Fisher fixed point $\left(0,0,\frac{\epsilon}{80}\right)$, a conformal QED $\times$ Wilson-Fisher fixed point $\left(\frac{3\epsilon}{8N_{f}},0,\frac{\epsilon}{80}\right)$, two pure $O(2)$ GNY fixed points with $e_{*}^{2}=0$ and $g_{*}^{2}\neq0,\lambda_{*}^{2}\neq0$, and two fixed points with all three couplings nonzero. The stable fixed-point is given by one of the latter two fixed points~\cite{scherer2016}:
\begin{eqnarray}
e_{*}^{2} & = & \frac{3}{8N_{f}}\epsilon+O\left(\epsilon^{2}\right),\label{QEDGNYFPe2}\\
g_{*}^{2} & = & \frac{9+2N_{f}}{4N_{f}\left(N_{f}+1\right)}\epsilon+O\left(\epsilon^{2}\right),\label{QEDGNYFPg2}\\
\lambda_{*}^{2} & = & \frac{Y-N_{f}-8}{160\left(N_{f}+1\right)}\epsilon+O\left(\epsilon^{2}\right)\label{QEDGNYFPl2},
\end{eqnarray}
where 
\begin{equation}
Y\equiv\sqrt{N_{f}^{2}+56N_{f}+424+\frac{810}{N_{f}}.}
\end{equation}
These three (squared) coupling constants are positive for all $N_{f}\geq1$,
including the cases $N_{f}=4,6,8$ applicable to the VBS phase
transition found numerically.

\subsection{Fate of $\mathbb{Z}_4$ anisotropy at criticality}
\label{sec:Z4anis}

\begin{figure}[t]
\includegraphics[width=0.8\columnwidth]{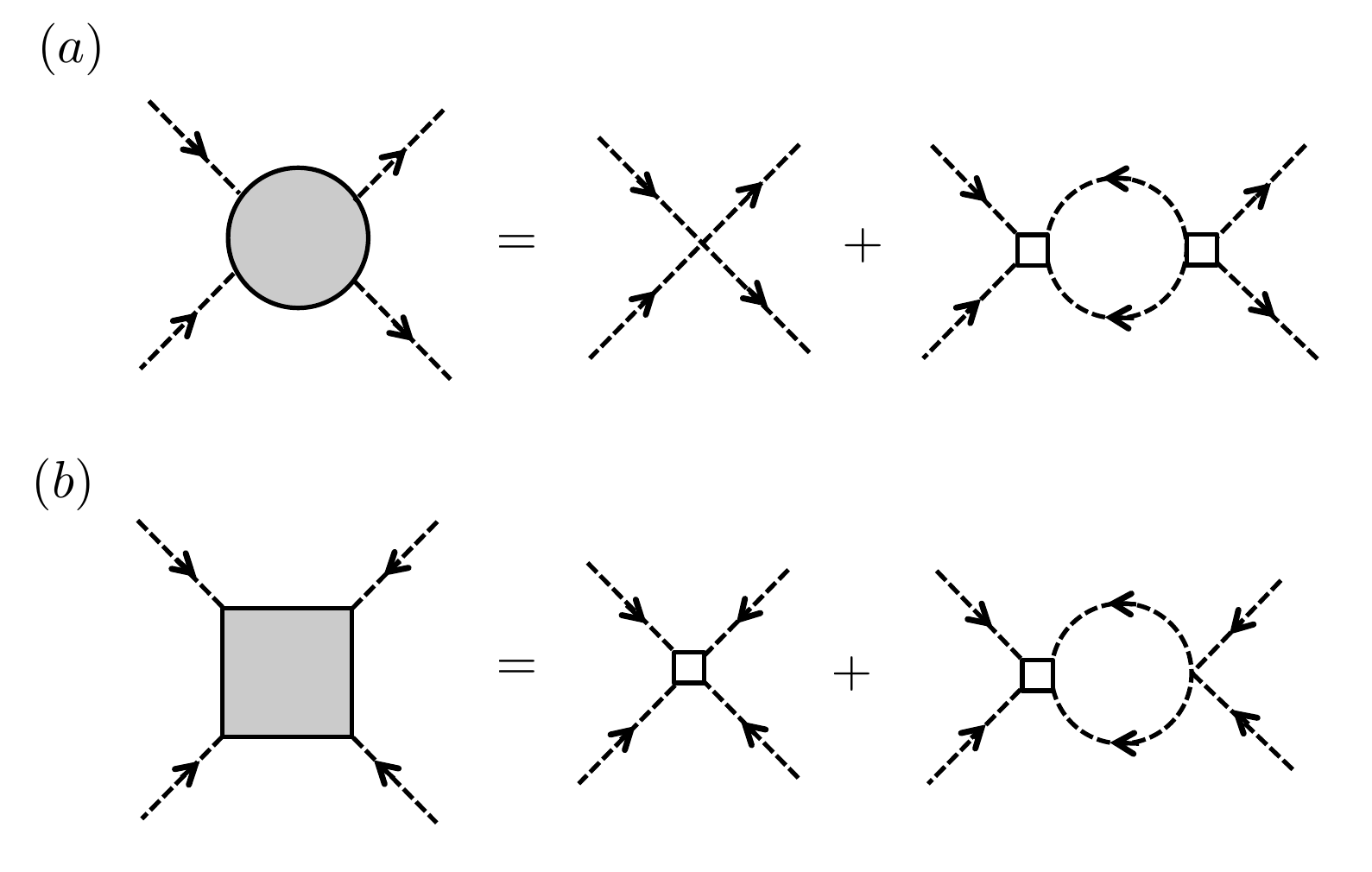}
\caption{One-loop renormalization of four-point vertices due to $\mathbb{Z}_4$ anisotropy (dashed lines: scalar field propagator, ordinary crossing: $|\phi|^4$ vertex, box: $\phi^4$ or $\phi^{*4}$ vertex). Renormalization of (a) the four-scalar coupling $\lambda^2$, (b) the anisotropy $b$.}
\label{fig:diagrams}
\end{figure}

It has so far only been shown that there exists a stable RG fixed point when the anisotropy term (\ref{Lanis}) is set to zero and the Lagrangian has an exact $U(1)$ symmetry. In reality, the long-wavelength theory will generically have $b\neq 0$ and the exact symmetry is only a discrete subgroup $\mathbb{Z}_4\subset U(1)$. For the chiral $O(2)$ QED$_3$-GNY model to describe the critical properties of the VBS transition, and assuming the anisotropy is small, we must show that the $U(1)$ symmetry is truly emergent, i.e., that the anisotropy is an irrelevant perturbation at the fixed point found in Sec.~\ref{sec:QCP}. For the Kekul\'e VBS transition on the honeycomb lattice in the presence of a dynamical gauge field, the $\mathbb{Z}_3$ anisotropy $\sim(\phi^3+\phi^{*3})$ is strongly relevant at small $\epsilon$ for finite $N_f$~\cite{scherer2016}, and its flow cannot be reliably controlled in the $\epsilon$ expansion. By contrast, the $\mathbb{Z}_4$ anisotropy (\ref{Lanis}) is marginal in four dimensions; its flow can thus be reliably controlled in the $\epsilon$ expansion. In the $\mathbb{Z}_3$ case, at one-loop order the anisotropy vertex contributes to the renormalization of the scalar-field two-point function (self-energy), three-point function (anisotropy vertex), and four-point function ($|\phi|^4$ vertex). By contrast, in the $\mathbb{Z}_4$ case the anisotropy does not contribute to the scalar-field self-energy at one-loop order, only to the scalar-field four-point functions (Fig.~\ref{fig:diagrams}). The couplings in the bare and renormalized anisotropy Lagrangians,
\begin{align}
\mathcal{L}_\text{anis.}^0&=b_0(\phi_0^4+\phi_0^{*4}),\\
\mathcal{L}^R_\text{anis.}&=Z_bb\mu^\epsilon(\phi^4+\phi^{*4}),
\end{align}
are related by $b=b_0\mu^{-\epsilon}Z_\phi^2Z_b^{-1}$. The beta function for $b$ is thus given by
\begin{align}
\beta_b=(-\epsilon+2\gamma_\phi-\gamma_b)b,
\end{align}
where $\gamma_b=d\ln Z_b/d\ln \mu$. Evaluating the diagrams in Fig.~\ref{fig:diagrams}, we find additional contributions to the one-loop renormalization constants $Z_{\lambda^2}$ and $Z_b$,
\begin{align}
Z_{\lambda^2}&\rightarrow Z_{\lambda^2}+\frac{576b^2\lambda^{-2}}{\epsilon},\\
Z_b&=1+\frac{96\lambda^2}{\epsilon},
\end{align}
which allows us to the find the corrected beta functions
\begin{align}
\beta_{\lambda^2}&\rightarrow\beta_{\lambda^2}+576b^2,\\
\beta_b&=(-\epsilon+4N_fg^2+96\lambda^2)b.
\end{align}
Since the anisotropy contribution to $\beta_{\lambda^2}$ is quadratic in $b$, the RG eigenvalue $y_b$ describing the flow of the anisotropy near the $O(2)$-symmetric QED$_3$-GNY fixed point is given simply by the negative of the slope of the UV beta function $\beta_b$ evaluated at the fixed point (\ref{QEDGNYFPe2})-(\ref{QEDGNYFPl2}),
\begin{align}\label{yb}
y_b=\epsilon-4N_fg_*^2-96\lambda_*^2,
\end{align}
such that $y_b>0$ denotes a relevant coupling. At the $O(2)$ Wilson-Fisher fixed point $e_*^2=g_*^2=0$ and $\lambda_*^2=\epsilon/80$, and one finds $y_b=-\epsilon/5$, in agreement with the analysis of pure $|\phi|^4$ theory perturbed by a $\mathbb{Z}_4$ anisotropy in Ref.~\cite{oshikawa2000}. At the QED$_3$-GNY fixed point (\ref{QEDGNYFPe2})-(\ref{QEDGNYFPl2}), we find
\begin{align}
y_b=-\frac{(2N_f+16+3Y)}{5(N_f+1)}\epsilon,
\end{align}
which is strictly negative for all $N_f$. Thus the $\mathbb{Z}_4$ anisotropy is irrelevant at small $\epsilon$ even for small $N_f$, in contrast with the $\mathbb{Z}_3$ case. This establishes the emergent $O(2)$ symmetry at the VBS QCP, and the fixed point discussed in Sec.~\ref{sec:QCP} is the true QCP. In remainder of Sec.~\ref{sec:epsilon} we compute its critical exponents at four-loop order in the $\epsilon$ expansion.

Before moving on to the critical properties of the $O(2)$ QED$_3$-GNY model, we observe that the calculation above can also address the issue of $\mathbb{Z}_4$ anisotropy at the semimetal-columnar VBS quantum phase transition of the $SU(4)$ Hubbard model on the $\pi$-flux square lattice~\cite{zhou2018}, which is described by the chiral $O(2)$ GNY model or ungauged NJL model supplemented by the anisotropy term (\ref{Lanis}). At one-loop order, the stable fixed point of the ungauged model is obtained by setting $e^2=0$ in the beta functions (\ref{eq:betag21}) and (\ref{eq:betal21}). We obtain
\begin{align}
(g_*^2)_\text{GNY}&=\frac{1}{2(N_f+1)}\epsilon,\\
(\lambda_*^2)_\text{GNY}&=\frac{\sqrt{N_f^2+38N_f+1}-N_f+1}{160(N_f+1)}\epsilon.
\end{align}
Substituting into Eq.~(\ref{yb}), we obtain
\begin{align}
(y_b)_\text{GNY}=-\frac{3\sqrt{N_f^2+38N_f+1}+2N_f-2}{5(N_f+1)}\epsilon,
\end{align}
which is strictly negative for all $N_f$. Thus the $\mathbb{Z}_4$ anisotropy is irrelevant at criticality also for the pure GNY model.

\subsection{Order-parameter anomalous dimension }

The order-parameter anomalous dimension $\eta_{\phi}$ characterizes the long-range power-law decay of the two-point function of the order parameter at criticality~\cite{QPT}. Here, this is the VBS correlation function,
\begin{align}\label{VBScorr}
\langle\mathcal{O}_\text{VBS}(r)\mathcal{O}_\text{VBS}(r')\rangle\sim\frac{1}{\left|r-r^{\prime}\right|^{1+\eta_{\phi}}}.
\end{align}
Microscopically, $\mathcal{O}_\text{VBS}(r)$ can be chosen as either component of the VBS order parameter $\b{V}=(V_x,V_y)$, where $V_x=(-1)^{x}\sum_{A}S_{A}(r)S_{A}(r+\hat{x})$ and $V_y=(-1)^{y}\sum_{A}S_{A}(r)S_{A}(r+\hat{y})$ correspond to columnar VBS order in the $x$ and $y$ directions, respectively. The $SU(N_{f})$ spin operator $S_{A}(r)$ is defined by
$S_{A}(r)=\sum_{ij}c_{ri}^{\dagger}T_{A}^{ij}c_{rj}$,
where $T_{A}$, $A=1,\dots,N_{f}^{2}-1$, are Hermitian generators of the $SU(N_{f})$ Lie group in the fundamental representation, and we choose the normalization $\tr T_AT_B=\delta_{AB}$. Note that by using the identity
\begin{align}
\sum_A T_A^{ij}T_A^{kl}=\delta^{il}\delta^{jk}-\frac{1}{N_f}\delta^{ij}\delta^{kl},
\end{align}
and the canonical anticommutation relations of fermion operators, one finds
\begin{align}
\sum_{ij}S_j^i(r)S_i^j(r')=\sum_A S_A(r)S_A(r'),
\end{align}
where $S_j^i(r)=c_{ri}^\dag c_{rj}-\frac{\delta_{ij}}{N_f}\sum_k c_{rk}^\dag c_{rk}$. Thus the $(\pi,0)$ and $(0,\pi)$ dimer operators $D^x_r=(-1)^x\sum_{ij}S_j^i(r)S_i^j(r+\hat{x})$ and $D^y_r=(-1)^y\sum_{ij}S_j^i(r)S_i^j(r+\hat{y})$ used in the QMC simulations~\cite{Meng2019,Meng2019b,janssen2020} coincide with the operators $V_x$ and $V_y$ defined above. In practice, one often computes equal-time correlation functions, such that $r$ and $r'$ in (\ref{VBScorr}) are spatial (lattice) coordinates with $|r-r'|\gg a$, $a$ being the lattice constant. As a consequence of Eq.~(\ref{VBScorr}), the anomalous dimension $\eta_\phi$ also appears in the finite-size analysis of the VBS structure factor. For instance, the $(\pi,0)$ VBS structure factor on an $L\times L$ lattice is given by
\begin{align}\label{SVBS}
S^{(\pi,0)}_\text{VBS}(L)&=\frac{1}{L^4}\sum_{rr'}\langle V_x(r)V_x(r')\rangle\nonumber\\
&\sim\frac{1}{L^4}\int d^2r\int d^2r'\frac{1}{|r-r'|^{1+\eta_\phi}}\sim L^{-(1+\eta_\phi)},
\end{align}
at criticality $J=J_c$, where we have approximated the lattice sum by a continuous integral, and cut the latter off at long distances by the system size and at short distances by the lattice constant $a$, here set to unity~\cite{sandvik2010}. Away from criticality $J\neq J_c$, one has
\begin{align}
S^{(\pi,0)}_\text{VBS}(L)\sim L^{-(1+\eta_\phi)}\mathcal{F}(L^{1/\nu}(J-J_c)),
\end{align}
where $\mathcal{F}$ is a universal scaling function.

In the field-theoretic approach,
$\eta_{\phi}$ is calculated by evaluating $\gamma_{\phi}$ at the QCP:
\begin{equation}
\eta_{\phi}=\gamma_{\phi}\left(e_{*}^{2},g_{*}^{2},\lambda_{*}^{2}\right).\label{eq:etaphi_eps}
\end{equation}
The anomalous dimensions can be expressed in a similar way as
the beta functions by a sum of contributions at a fixed loop order:
\begin{equation}
\gamma_{X}=\gamma_{X}^{\left(\text{1L}\right)}+\gamma_{X}^{\left(\text{2L}\right)}+\gamma_{X}^{\left(\text{3L}\right)}+\gamma_{X}^{\left(\text{4L}\right)}.
\end{equation}
For the scalar anomalous dimension, the contributions at each order
are given by 
\begin{eqnarray}
\gamma_{\phi}^{\left(1\text{L}\right)} & = & 2N_{f}g^{2},\\
\gamma_{\phi}^{\left(2\text{L}\right)} & = & 10N_{f}e^{2}g^{2}-3N_{f}g^{4}+128\lambda^{4},\\
\gamma_{\phi}^{\left(3\text{L}\right)} & = & -\frac{1}{2}N_{f}e^{4}g^{2}\left(32N_{f}-144\zeta_{3}+119\right)\nonumber\\
 &  & -\frac{1}{2}N_{f}e^{2}g^{4}\left(48\zeta_{3}-5\right)+80N_{f}g^{4}\lambda^{2}\nonumber \\
 &  & +\frac{1}{8}N_{f}g^{6}\left(64N_{f}+48\zeta_{3}-27\right)\nonumber \\
 &  & -480N_{f}g^{2}\lambda^{4}-2560\lambda^{6}.
\end{eqnarray}
At one-loop order, we obtain
\begin{equation}
\eta_{\phi}=\frac{2N_{f}+9}{2\left(N_{f}+1\right)}\epsilon+O\left(\epsilon^{2}\right),
\end{equation}
in agreement with Ref.~\cite{janssen2020}. The four-loop result is presented in Ref.~\cite{SuppMat}. 
The four-loop results to $O\left(\epsilon^{4}\right)$ for $N_{f}=4,6,8$ are respectively given by 
\begin{align}
\eta_{\phi} & \approx  1.7\epsilon+0.05330\epsilon^{2}+0.9040\epsilon^{3}-3.455\epsilon^{4},\\
\eta_{\phi} & \approx  1.5\epsilon-0.02886\epsilon^{2}+0.3396\epsilon^{3}-1.075\epsilon^{4},\\
\eta_{\phi} & \approx  1.389\epsilon-0.04893\epsilon^{2}+0.1597\epsilon^{3}-0.4775\epsilon^{4}.
\end{align}

\begin{figure}[t]
\centering\includegraphics[width=\columnwidth]{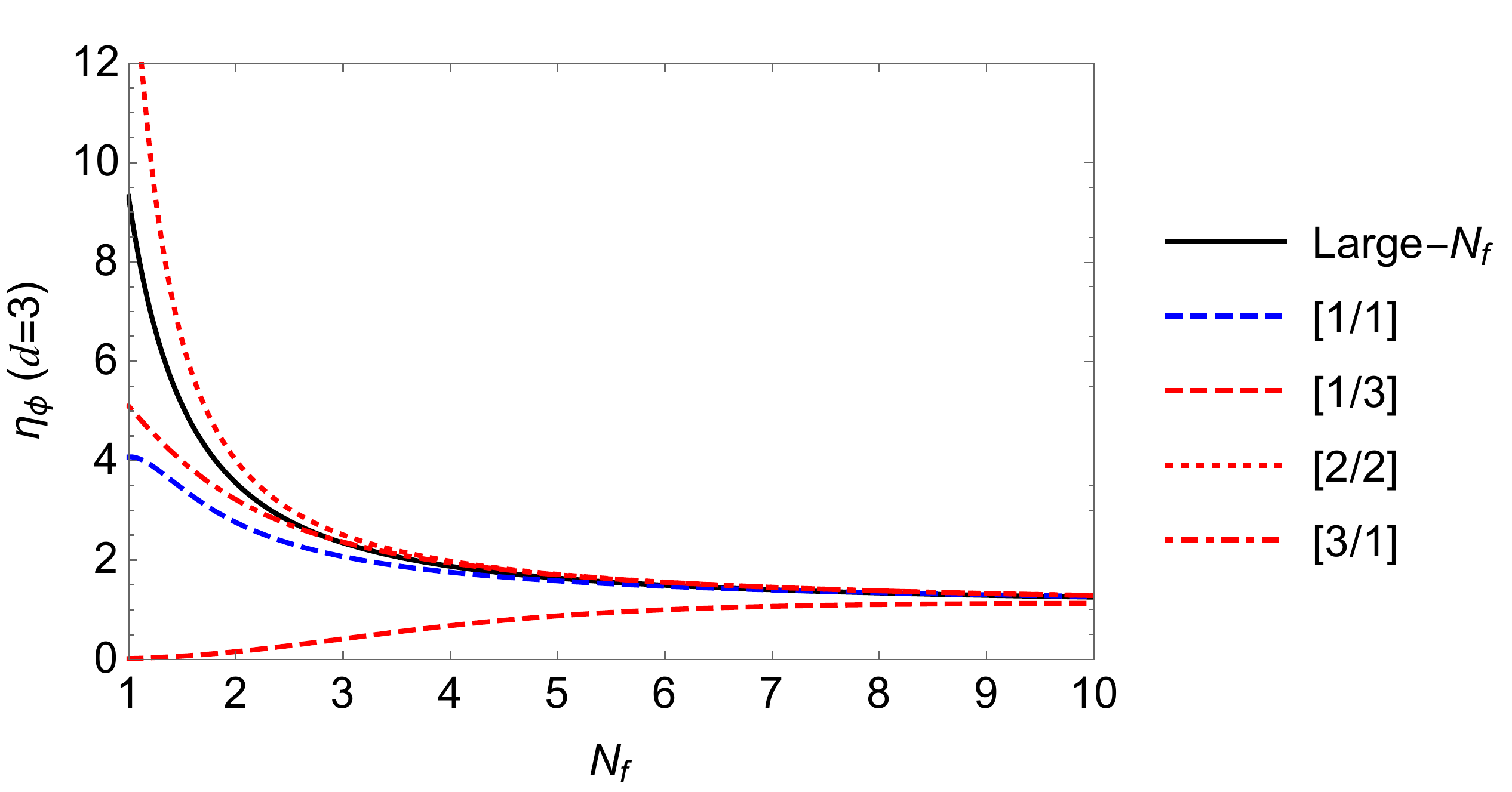}
\caption{Pad\'e approximants for $\eta_{\phi}$ as a function of $N_{f}$ at two (blue) and four-loop (red) orders. The large-$N_f$ result (\ref{eq:etaphi_largeN}) is shown in black.}
\label{fig:etaphi_d3}
\end{figure}
Both the $\epsilon$-expansion and the large-$N_{f}$ expansion (see Sec.~\ref{sec:largeN}) generate asymptotic series that have zero radius of convergence. In order to
extract physical results from finite-order $\epsilon$-expansion
expressions derived perturbatively, resummation procedures
must be implemented~\cite{KleinertPhi4}. Two standard resummation techniques are Pad\'e approximants, discussed below, and the Pad\'e-Borel transformation, reviewed in Appendix~\ref{app:QEDGN}. For a given loop order $L$, the (one-sided) Pad\'e approximants are defined by
\begin{equation}
\left[m/n\right]\left(\epsilon\right)=\frac{\sum_{i=0}^{m}a_{i}\epsilon^{i}}{1+\sum_{j=1}^{n}b_{j}\epsilon^{j}}.
\end{equation}
Here, $m$ and $n$ are two positive integers satisfying $m+n=L$. The coefficients $a_{i}$ and $b_{j}$ are determined such that expanding the above function in powers of $\epsilon$ to $O(\epsilon^{L})$ reproduces the $\epsilon$-expansion results. In Fig.~\ref{fig:etaphi_d3} we plot Pad\'e approximants (colored lines) in $d=3$ at two and four-loop orders; three-loop approximants turn out to have poles in the extrapolation region $\epsilon\in[0,1]$ for certain values of $N_f$ in the range considered, and are thus excluded from the plot. Apart from the [1/3] approximant, a good convergence of the approximants with increasing loop order is found for the values of $N_f=4,6,8$ studied in QMC. Numerical values of Pad\'e and Pad\'e-Borel approximants for $\eta_\phi$ for $N_f=4,6,8$ are given in Appendix~\ref{app:QEDGNepsilon} (Tables \ref{tab:PadeN4_EpsilonExp}, \ref{tab:PadeN6_EpsilonExp}, and \ref{tab:PadeN8_EpsilonExp}, respectively).

\subsection{Correlation length exponent}

The correlation length exponent $\nu$ governs the divergence of the zero-temperature correlation length as the QCP is approached, i.e., as the scalar mass squared $m^2$ is tuned to zero. The anomalous dimension for the scalar mass squared is defined by $\gamma_{m^{2}}=\gamma_{\phi^{2}}-\gamma_{\phi}$, where $\gamma_{\phi}$, the anomalous dimension of the order-parameter field $\phi$, has already been computed in the previous section. The beta function for $m^{2}$ is given by
\begin{equation}
\mu\frac{dm^{2}}{d\mu}=-\left(2+\gamma_{m^{2}}\right)m^{2}.
\end{equation}
At the QCP, the correlation length exponent $\nu$ is related to the
anomalous dimension $\gamma_{m^{2}}$ by 
\begin{equation}
1/\nu=2+\gamma_{m^{2}}\left(e_{*}^{2},g_{*}^{2},\lambda_{*}^{2}\right).\label{eq:nuinv_eps}
\end{equation}
The contributions to $\gamma_{m^{2}}$, up to three-loop order, are
given by 
\begin{eqnarray}
\gamma_{m^{2}}^{\left(1\text{L}\right)} & = & -2N_{f}g^{2}-32\lambda^{2},\\
\gamma_{m^{2}}^{\left(2\text{L}\right)} & = & 3N_{f}g^{4}-10N_{f}e^{2}g^{2}
  +64N_{f}g^{2}\lambda^{2}+640\lambda^{4},\\
\gamma_{m^{2}}^{\left(3\text{L}\right)} & = & \frac{1}{2}N_{f}e^{4}g^{2}\left(32N_{f}-144\zeta_{3}+119\right)\nonumber\\
 &  & -\frac{1}{8}N_{f}g^{6}\left(256N_{f}-379+240\zeta_{3}\right)\nonumber\\
 &  & +8N_{f}g^{4}\lambda^{2}\left(12N_{f}-48\zeta_{3}-35\right)\nonumber\\
 &  & +\frac{1}{2}N_{f}e^{2}g^{4}\left(240\zeta_{3}-149\right)\nonumber\\
 &  & -48N_{f}e^{2}g^{2}\lambda^{2}\left(16\zeta_{3}-17\right)\nonumber\\
 &  & -1056N_{f}g^{2}\lambda^{4}-72192\lambda^{6}.
\end{eqnarray}
At one-loop order, we obtain
\begin{equation}
1/\nu=2-\frac{8N_{f}+29+2Y}{10\left(N_{f}+1\right)}\epsilon+O\left(\epsilon^{2}\right),
\end{equation}
in agreement with Ref.~\cite{janssen2020}. The four-loop order result is presented in Ref.~\cite{SuppMat}. The four-loop
order results to $O\left(\epsilon^{4}\right)$ for $N_{f}=4,6,8$ are respectively given by 
\begin{align}
1/\nu & \approx  2-2.397\epsilon+1.484\epsilon^{2}-4.376\epsilon^{3}+16.46\epsilon^{4},\\
1/\nu & \approx  2-1.972\epsilon+0.8886\epsilon^{2}-1.562\epsilon^{3}+4.193\epsilon^{4},\\
1/\nu & \approx  2-1.749\epsilon+0.6315\epsilon^{2}-0.7591\epsilon^{3}+1.639\epsilon^{4}.
\end{align}

\begin{figure}[t]
\centering\includegraphics[width=\columnwidth]{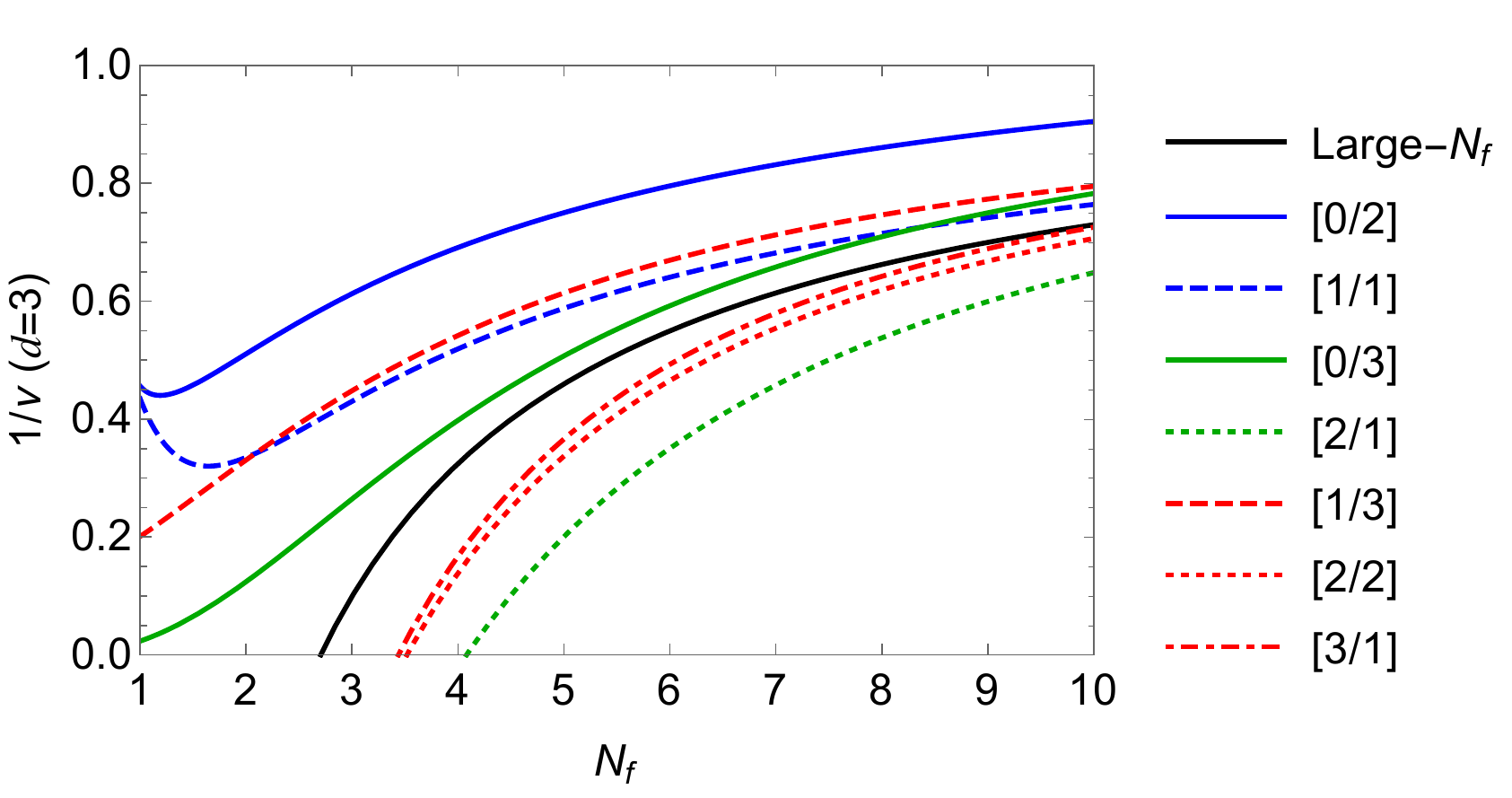}
\caption{Pad\'e approximants for $1/\nu$ as a function of $N_{f}$ at two (blue), three (green), and four-loop (red) orders. The large-$N_f$ result (\ref{eq:nuinv_largeN}) is shown in black.}
\label{fig:nuInv_d3}
\end{figure}

Pad\'e approximants for $1/\nu$ in $d=3$ up to four-loop order are plotted using colored lines in Fig.~\ref{fig:nuInv_d3}. Relatively poor convergence with increasing loop order is found, and it is difficult to obtain reliable estimates even for $N_f=4,6,8$. Numerical values of Pad\'e and Pad\'e-Borel approximants for $1/\nu$ for those values of $N_f$ are given in Appendix~\ref{app:QEDGNepsilon} (Tables \ref{tab:PadeN4_EpsilonExp}, \ref{tab:PadeN6_EpsilonExp}, and \ref{tab:PadeN8_EpsilonExp}, respectively).

\subsection{Fermion bilinears and CDW exponent}
\label{sec:FermionBilinearCDW}

Apart from the scaling dimension $\Delta_\text{VBS}=\Delta_\phi=(1+\eta_\phi)/2$, which controls VBS two-point correlations at criticality [see Eqs.~(\ref{VBScorr}) and (\ref{SVBS})], the scaling dimension
of other gauge-invariant local operators can be computed. Other microscopic gauge-invariant local observables for model (\ref{eq:Hamiltonian}) and accessible to QMC simulations include the staggered density or CDW operator $\mathcal{O}_{\text{CDW}}(r)=(-1)^{x+y}\sum_{i}c_{ri}^{\dagger}c_{ri}$,
the staggered $SU(N_{f})$ spin $\mathcal{O}_\text{AF}(r)=(-1)^{x+y}S_{A}(r)$, and a QAH mass operator $\mathcal{O}_{\text{QAH}}(r)$~\cite{boyack2019b}. These operators also exhibit universal power-law correlations $\langle\mathcal{O}(r)\mathcal{O}(r')\rangle\sim|r-r'|^{-2\Delta_\mathcal{O}}$ at criticality, which correspond to non-diverging static susceptibilities $\chi_{\mathcal{O}}(\b{q})\sim\left|\boldsymbol{q}\right|^{2\Delta_{\mathcal{O}}-3}$, with $\Delta_\mathcal{O}>3/2$. (By contrast, the static VBS susceptibility diverges as $\chi_\text{VBS}(\b{q})\sim|\b{q}|^{2\Delta_\phi-3}\sim|\b{q}|^{-(2-\eta_\phi)}$, due to critical fluctuations of the VBS order parameter.)  The microscopic observables above correspond in the long-wavelength, low-energy effective field theory to Lorentz-invariant, gauge-invariant fermion bilinears~\cite{Zerf2019,boyack2019b}. In terms of the fermion fields $\Psi,\overline{\Psi}$ in the $O(2)$ QED$_3$-GNY model (\ref{eq:XYQEDGN_Model}), the identification is
\begin{eqnarray}
\mathcal{O}_{\text{CDW}} & \sim & \overline{\Psi}\Psi,\label{OCDWO2QEDGN}\\
\mathcal{O}_\text{AF} & \sim & \overline{\Psi}T_{A}\Psi,\label{OAFO2QEDGN}\\
\mathcal{O}_{\text{QAH}} & \sim & i\overline{\Psi}\Gamma_{3}\Gamma_{5}\Psi,\label{OQAHO2QEDGN}
\end{eqnarray}
while in the gauged NJL formulation (\ref{eq:Lagrangian_NJL}) with the fields $\psi,\overline{\psi}$, these bilinears are 
\begin{eqnarray}
\mathcal{O}_{\text{CDW}} & \sim & -i\overline{\psi}\gamma_{3}\psi,\label{OCDWNJL}\\
\mathcal{O}_\text{AF} & \sim & -i\overline{\psi}T_{A}\gamma_{3}\psi,\label{OAFNJL}\\
\mathcal{O}_{\text{QAH}} & \sim & i\overline{\psi}\gamma_{3}\gamma_{5}\psi.\label{OQAHNJL}
\end{eqnarray}
In both sets of equations a sum over repeated flavor indices is understood. Since our four-loop analysis is based on a $d$-dimensional representation
of the gauged NJL model, the appearance of the $\gamma_{3}$ matrix makes the dimensional continuation of the above fermion bilinears while preserving their Lorentz invariance appear intractable at
first sight (see Ref.~\cite{Zerf2019} for a discussion of related issues in the context of the Ising QED$_3$-GNY model). To overcome this issue, in our diagrammatic calculations we introduce a formal object $\gamma_3'=(\gamma_3')^\dag$ which squares to the identity, is traceless, and naively anticommutes with all gamma matrices: $\{\gamma_3',\gamma_5\}=0$ and $\{\gamma_3',\gamma_\mu\}=0$ for $\mu=0,1,\ldots,d-1$ in general $d$ dimensions. This object obeys the same properties as $\gamma_3$ in $d=3$ dimensions, and is thus a suitable replacement for $\gamma_3$ in the bilinears (\ref{OCDWNJL})-(\ref{OQAHNJL}) for general $d$ calculations. As an example application of this procedure, we calculate the scaling dimension of the simplest bilinear of this type, the CDW operator (\ref{OCDWNJL}), thus defined as $\Delta_\text{CDW}\equiv\Delta_{i\overline{\psi}\gamma_3'\psi}$ in general $d$. The corresponding anomalous dimension is given by
\begin{figure}[t]
\centering\includegraphics[width=\columnwidth]{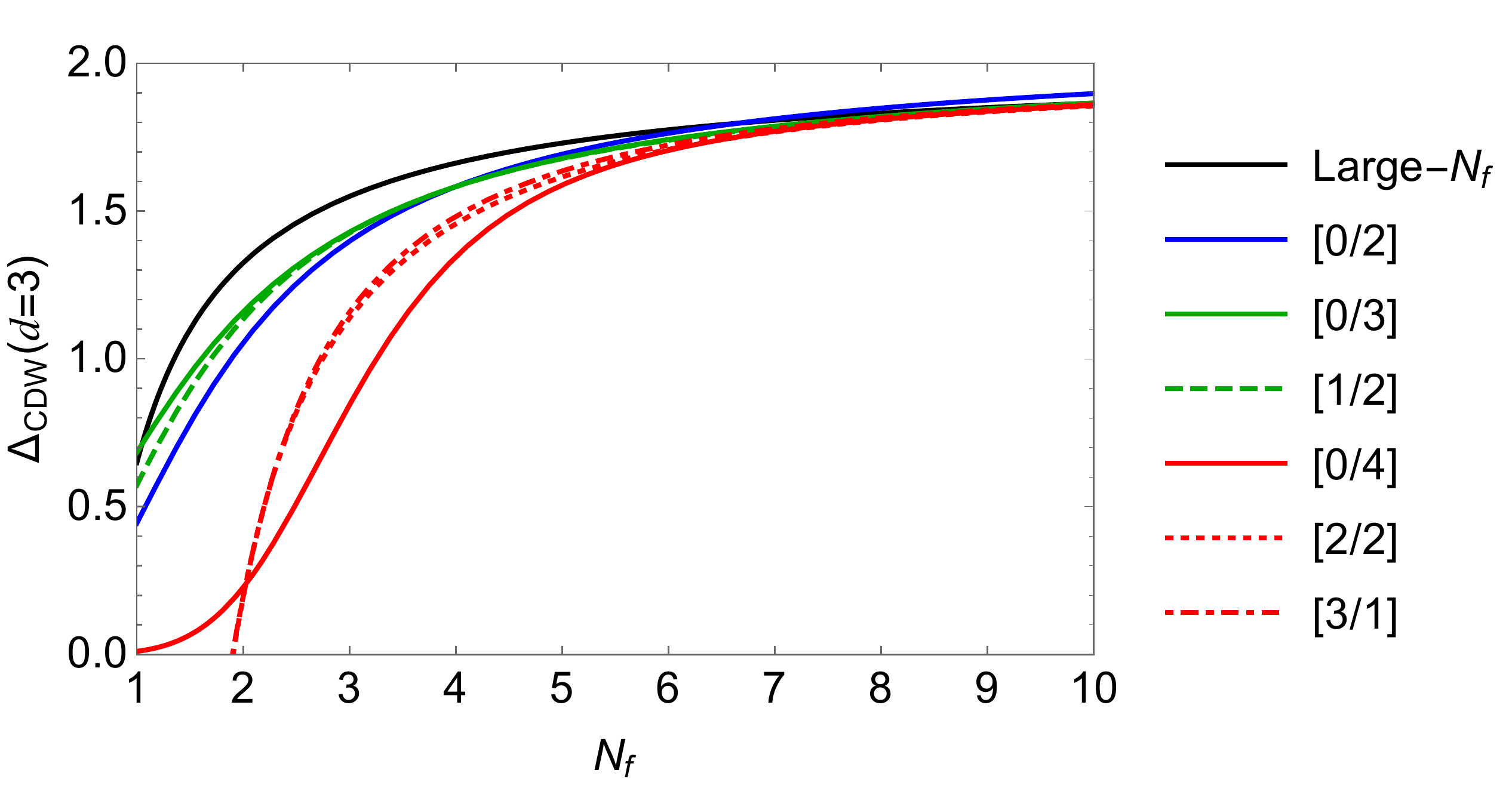}
\caption{Pad\'e approximants for $\Delta_\text{CDW}$ as a function of $N_{f}$ for $d=3$ at two (blue), three (green), and four-loop (red) orders. The large-$N_f$ result (\ref{eq:delCDW_largeN}) is shown in black.}
\label{fig:delCDW_d3}
\end{figure}
\begin{eqnarray}
\gamma_\text{CDW}^{\left(1\text{L}\right)} & = & 6e^{2}+g^{2},\\
\gamma_\text{CDW}^{\left(2\text{L}\right)} & = & -\frac{1}{3}e^4 \left(20N_{f}-9\right)+8 e^2 g^2\nonumber\\
 &  &-\frac{1}{4} g^4 (2 N_{f}-1),\\
\gamma_\text{CDW}^{\left(3\text{L}\right)} & = & -\frac{1}{27}e^{6}\left[4N_{f}\left(70N_{f}-648\zeta_{3}+621\right)-3483)\right]\nonumber\\
 &  & +\frac{1}{4}e^{4}g^{2}\left(32N_{f}-720\zeta_{3}+13\right)\nonumber\\
 &  & +\frac{1}{4}e^{2}g^{4}\left[N_{f}\left(48\zeta_{3}-19\right)-96\zeta_{3}+72\right]\nonumber \\
 &  & -\frac{1}{16}g^{6}\left[2N_{f}\left(10N_{f}-31\right)+48\zeta_{3}-57\right]\nonumber \\
 &  & +16g^{4}\lambda^{2}-28g^{2}\lambda^{4},
\end{eqnarray}
at three-loop order, with the four-loop contribution given in Ref.~\cite{SuppMat}.
At one-loop order, we obtain
\begin{equation}
\Delta_\text{CDW}=3-\frac{4N_{f}^{2}+15N_{f}+18}{4N_{f}\left(N_{f}+1\right)}\epsilon+O\left(\epsilon^{2}\right).
\end{equation}
The four-loop order results for $N_{f}=4,6,8$ are respectively given by 
\begin{align}
\Delta_\text{CDW} & =  3-1.775\epsilon+0.1394\epsilon^{2}+0.4558\epsilon^{3}-1.317\epsilon^{4},\\
\Delta_\text{CDW} & =  3-1.500\epsilon+0.1453\epsilon^{2}+0.1677\epsilon^{3}-0.1937\epsilon^{4},\\
\Delta_\text{CDW} & =  3-1.368\epsilon+0.1223\epsilon^{2}+0.09757\epsilon^{3}-0.05859\epsilon^{4}.
\end{align}

We plot the corresponding Pad\'e approximants as colored lines in Fig.~\ref{fig:delCDW_d3}. Except for the four-loop approximants, all approximants agree quite well for $N_f=4,6,8$. Numerical values of the approximants for the latter are given in Appendix~\ref{app:QEDGNepsilon} (Tables \ref{tab:PadeN4_EpsilonExp}-\ref{tab:PadeN8_EpsilonExp}).

\subsection{Kekul\'e VBS transition on the honeycomb lattice: CDW exponent}
\label{sec:Kekule_eps}

\begin{figure}[t]
\centering\includegraphics[width=\columnwidth]{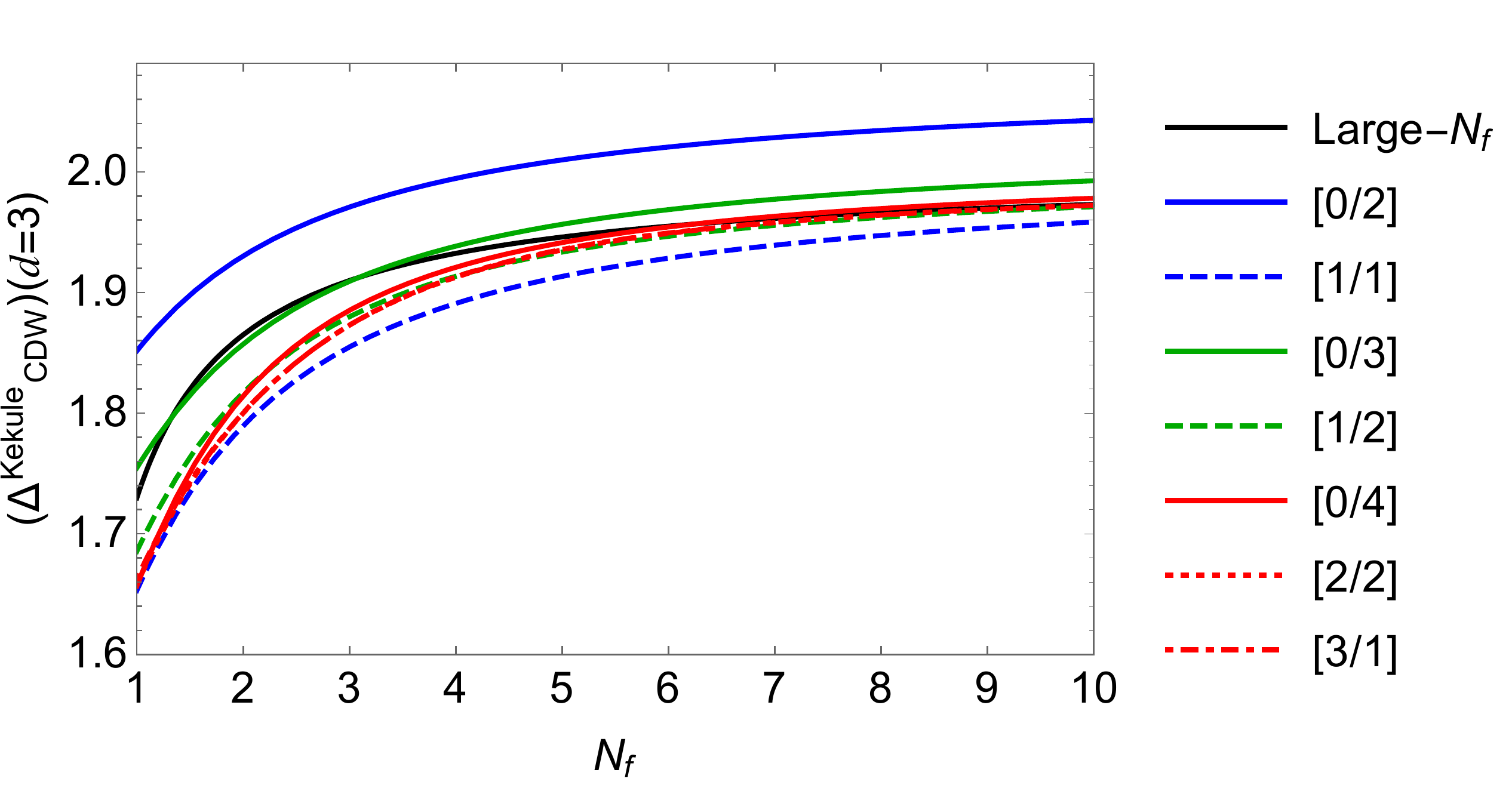}
\caption{Pad\'e approximants for $\Delta_\text{CDW}$ in the chiral $O(2)$ GNY model as a function of $N_{f}$ for $d=3$ at two (blue), three (green), and four-loop (red) orders. The large-$N_f$ result (\ref{CDWKekuleLargeN}) is shown in black.}
\label{fig:delCDWXY_d3}
\end{figure}

The CDW operator (\ref{OCDWNJL}) also corresponds to the long-wavelength limit of the staggered density (or Semenoff mass~\cite{semenoff1984}) on the honeycomb lattice. Evaluating the anomalous dimension $\gamma_\text{CDW}$ computed above at the stable fixed point of the chiral $O(2)$ GNY model or the ungauged NJL model, i.e., setting the gauge coupling to zero in the Lagrangian (\ref{eq:Lagrangian_NJL}), one can determine the universal exponent $\Delta_\text{CDW}^\text{Kekul\'e}$ characterizing the decay of CDW correlations at the QCP of the Kekul\'e VBS transition on the honeycomb lattice~\cite{lang2013,zhou2016,li2017b,li2020}. We assume here that $N_f$ is sufficiently large that the $\mathbb{Z}_3$ anisotropy is irrelevant at the QCP. At one-loop order, we obtain
\begin{align}
\Delta_\text{CDW}^\text{Kekul\'e} = 3-\frac{2N_{f}+3}{2(N_{f}+1)}\epsilon+O(\epsilon^2).
\end{align}
Evaluating the four-loop expressions~\cite{SuppMat} numerically for $N_f=2,3,4$, respectively, we obtain
\begin{align}
\Delta_\text{CDW}^\text{Kekul\'e} & =  3 -1.167 \epsilon-0.04222 \epsilon^2+0.02605 \epsilon^3-0.04662 \epsilon^4,\\
\Delta_\text{CDW}^\text{Kekul\'e} & =  3 -1.125 \epsilon -0.01980 \epsilon^2+0.02584 \epsilon^3-0.01168 \epsilon^4,\\
\Delta_\text{CDW}^\text{Kekul\'e} & =  3 -1.100 \epsilon-0.009 \epsilon^2+0.02384 \epsilon^3-0.001980 \epsilon^4.
\end{align}

The corresponding Pad\'e approximants are plotted as colored lines in Fig.~\ref{fig:delCDWXY_d3}. An excellent convergence with increasing loop order is found; in particular, all four-loop approximants agree very closely. Numerical values of the approximants for select values of $N_f$ are found in Appendix~\ref{app:GN}, Table~\ref{tab:XYPade_EpsilonExp}.

\section{Large-$N_{f}$ Expansion}
\label{sec:largeN}

The previous section is based on a perturbative analysis of the chiral $O(2)$ QED-GNY model (gauged NJL model) in $d=4-\epsilon$ spacetime dimensions up to $O(\epsilon^{4})$. Since the physical dimension of interest is $d=3$, it is pertinent to consider other approximation techniques that allow complementary aspects of the critical point to be illuminated. One such approach is the large-$N_{f}$ expansion, reviewed in Ref.~\cite{gracey2018review}, where one formally considers a large and arbitrary number of fermion flavors $N_{f}$ and constructs an expansion in powers of $1/N_{f}$.
A large-$N_{f}$ analysis for the chiral Ising QED-GNY theory was performed to $O(1/N_{f}^2)$, in $2<d<4$ spacetime dimensions, in Ref.~\cite{gracey2018} using the large-$N$ critical point formalism. 
A large-$N_{f}$ analysis of the same model, to $O(1/N_{f})$ but for fixed $d=3$, was performed in Refs.~\cite{Boyack2019,benvenuti2019}, and it was noted that in fixed $d=3$ certain critical exponents have contributions that are not captured in the continuous $2<d<4$ analysis.
For the chiral $O(3)$ QED$_{3}$-GNY theory, the large-$N_{f}$ analysis was performed in Ref.~\cite{Zerf2019}.

The chiral $O(2)$ QED$_3$-GNY model considered here was first studied in the large-$N_f$ expansion in Refs.~\cite{Gracey1993a} and \cite{boyack2019b}. In Ref.~\cite{Gracey1993a}, the critical exponents $\eta_\phi$ and $\nu$ were computed to $O(1/N_f)$ in $2<d<4$, while in Ref.~\cite{boyack2019b} the scaling dimensions of the CDW, AF, and QAH bilinears (\ref{OCDWO2QEDGN})-(\ref{OQAHO2QEDGN}) were computed to $O(1/N_f)$ in fixed $d=3$. Following the method used in Ref.~\cite{Gracey1993a}, here we expand upon these previous studies by computing $\eta_\phi$ to $O(1/N_f^2)$ and $\Delta_\text{CDW}$ to $O(1/N_f)$, in general $2<d<4$. We also establish consistency between the results of the $\epsilon$-expansion (Sec.~\ref{sec:epsilon}) and those of the large-$N_f$ expansion, by verifying that all exponents computed by both methods agree to $O(\epsilon^4,1/N_f^p)$, with $p=1$ or $2$ the order at which a given quantity is known in the large-$N_f$ expansion. This constitutes a strong check on both the $\epsilon$-expansion and large-$N_f$ expansion results.

\subsection{Critical exponents}

A critical exponent $x$ can be expanded in a series of the form $x=\sum\limits_{i=1}^{\infty}x_{i}/N^{i}_{f}$.  To first order in $1/N_{f}$, the pertinent quantities to compute here are the fermion anomalous dimension $\eta_{1}$ and the fermion-scalar vertex anomalous dimension $\chi_{\phi,1}$. As a result of the Ward-Takahashi identity, the fermion-gauge vertex obeys $\chi_{A,1}=-\eta_{1}$; thus, it is not an independent quantity. The fermion anomalous dimension is a gauge-dependent quantity and throughout this paper we consider the Landau gauge (with gauge fixing parameter $\xi=0$). To determine $1/\nu$ the exponent $\lambda$ (defined in Ref.~\cite{Gracey1993a}) must be computed. 
The results for these quantities at $O(1/N_{f}$) have already been computed~\cite{Gracey1993a}, and are reproduced here for convenience:
\begin{eqnarray}
\eta_{1} & = & -\frac{\left(4\mu^{3}-8\mu^{2}+\mu+2\right)}{4\left(\mu-1\right)}\frac{\Gamma\left(2\mu-1\right)}{\mu\Gamma\left(1-\mu\right)\Gamma\left(\mu\right)^{3}},\\
\chi_{\phi,1} & = & -\frac{\mu\left(2\mu-1\right)^{2}}{\left(4\mu^{3}-8\mu^{2}+\mu+2\right)}\eta_{1},\\
\lambda_{1} & = & -\frac{1}{2}\frac{\left(2\mu-1\right)\left(2\mu^{2}-2\mu+1\right)\Gamma\left(2\mu-1\right)}{\Gamma(2-\mu)\mu\Gamma\left(\mu\right)^{3}}.
\end{eqnarray}
Here $\Gamma(z)$ denotes Euler's gamma function and $\mu=d/2$. To $O(1/N_{f}^2)$, $\eta_{2}$ and $\chi_{\phi,2}$ are given by
\begin{widetext}
\begin{eqnarray}
\eta_{2} & = & \left[-8\left(2\mu^{2}-1\right)\left(\mu-1\right)^{2}\hat{\Psi}\left(\mu\right)+3\mu\left(2\mu-1\right)\left(4\mu^{3}-8\mu^{2}+\mu+2\right)\left(\mu-1\right)\left(\psi^{\prime}\left(\mu-1\right)-\psi^{\prime}\left(1\right)\right)\right.\nonumber \\
 &  & \left.+\frac{16\mu^{7}-112\mu^{6}+240\mu^{5}-184\mu^{4}+15\mu^{3}+46\mu^{2}-20\mu+2}{\mu\left(\mu-1\right)}\right]\frac{\eta_{1}^{2}}{\left(4\mu^{3}-8\mu^{2}+\mu+2\right)^{2}},\\
\chi_{\phi,2} & = & -\left[3\left(\mu-1\right)\mu^{2}\left(4\mu+1\right)\left(2\mu-1\right){}^{2}\left(\psi^{\prime}\left(\mu-1\right)-\psi^{\prime}\left(1\right)\right)\right.\nonumber \\
 &  & \left.+\frac{48\mu^{7}-184\mu^{6}+204\mu^{5}-30\mu^{4}-82\mu^{3}+31\mu^{2}+8\mu-2}{\mu-1}\right]\frac{\eta_{1}^{2}}{\left(4\mu^{3}-8\mu^{2}+\mu+2\right)^{2}},
\end{eqnarray}
\end{widetext}
where $\psi\left(z\right)\equiv\Gamma^{\prime}\left(z\right)/\Gamma\left(z\right)$ and 
\begin{equation}
\hat{\Psi}\left(\mu\right)=\psi\left(2\mu-1\right)-\psi\left(1\right)+\psi\left(1-\mu\right)-\psi\left(\mu-1\right).
\end{equation}
The scalar anomalous dimension is then determined via $\eta_{\phi}=2(2-\mu-(\eta+\chi_{\phi}))$, and the inverse correlation exponent is obtained from $1/\nu=2(\mu-1+\lambda)$. 
Expanding the expressions for $\eta_{\phi}$ and $1/\nu$ in $d=4-\epsilon$ up to $O(\epsilon^{4})$, we find agreement with the counterpart expressions in Eqs.~(\ref{eq:etaphi_eps}) and (\ref{eq:nuinv_eps}) respectively, when the latter are expanded in powers of $1/N_{f}$ to $O(1/N^{2}_f)$ and $O(1/N_{f})$ respectively. This agreement is an important verification of the validity of our results.  In fixed $d=3$, the large-$N_{f}$ expressions reduce to 
\begin{eqnarray}
\left.\eta_{\phi}\right|_{d=3} & = & 1+\frac{56}{3\pi^{2}N_{f}}+\frac{3168\pi^{2}-14368}{27\pi^{4}N_{f}^{2}}.\label{eq:etaphi_largeN}\\
\left.\frac{1}{\nu}\right|_{d=3} & = & 1-\frac{80}{3\pi^{2}N_{f}}. \label{eq:nuinv_largeN}
\end{eqnarray}
The results for $\eta_{\phi}$ and $1/\nu$ at $O(1/N_{f})$ agree with Ref.~\cite{Gracey1993a}. 
Note that, for the chiral $O(2)$ QED$_{3}$-GNY model, in fixed $d=3$ spacetime dimensions, there are no additional contributions at $O(1/N_{f}$) arising from Aslamazov-Larkin diagrams~\cite{Boyack2019},
and the above result agrees to $O(1/N_{f}$) with Ref.~\cite{boyack2019b}.

The large-$N_{f}$ results are plotted in Fig.~\ref{fig:etaphi_d3} ($\eta_\phi$) and Fig.~\ref{fig:nuInv_d3} ($1/\nu$) alongside the Pad\'e approximants for the $\epsilon$-expansion results. Excellent agreement with the $\epsilon$-expansion approximants is found for $\eta_\phi$ for $N_f\gtrsim 4$, while the large spread of values of the $\epsilon$-expansion approximants for $1/\nu$ prevents a meaningful comparison with the large-$N_f$ result. In Appendix~\ref{app:QED3GNlargeNf}, we also resum the large-$N_f$ results for $\eta_\phi$ and $1/\nu$ at $N_f=4$ (Table~\ref{tab:PadeN4_LargeNExp}), $N_f=6$ (Table~\ref{tab:PadeN6_LargeNExp}), and $N_f=8$ (Table~\ref{tab:PadeN8_LargeNExp}), treating $1/N_f$ as a small parameter and using Pad\'e and Pad\'e-Borel resummation.

\subsection{CDW exponent}

As discussed in Sec.~\ref{sec:FermionBilinearCDW}, in general $2<d<4$ dimensions a suitable definition of the CDW exponent is as the scaling dimension of the $i\overline{\psi}\gamma_3'\psi$ fermion bilinear in the gauged NJL model. In the large-$N_f$ formalism, this exponent is given by
\begin{equation}\label{DCDWlargeN}
\Delta_\text{CDW}\equiv\Delta_{i\overline{\psi}\gamma_{3}^{\prime}\psi}=2\mu-1+\eta_{\mathcal{O},3^{\prime}},
\end{equation}
where the parameter $\eta_{\mathcal{O},3^{\prime}}$ is given by
\begin{equation}
\eta_{\mathcal{O},3^{\prime}}=-\frac{2\left(4\mu^{2}-2\mu-1\right)}{\left(4\mu^{3}-8\mu^{2}+\mu+2\right)}\eta_{1}.
\end{equation}
Again, when Eq.~(\ref{DCDWlargeN}) is expanded in $d=4-\epsilon$ up to $O(\epsilon^{4})$, we find agreement with the counterpart expression computed at four-loop order in Sec.~\ref{sec:FermionBilinearCDW}, when the latter is expanded in powers of $1/N_{f}$ to $O(1/N_{f}$).
In fixed $d=3$ the result is 
\begin{eqnarray}
\left.\Delta_\text{CDW}\right|_{d=3} & = & 2-\frac{40}{3\pi^{2}N_{f}}.\label{eq:delCDW_largeN}
\end{eqnarray}
This agrees with Ref.~\cite{boyack2019b}, which computed $\Delta_{\overline{\Psi}\Psi}=\Delta_\text{CDW}$ in the $O(2)$ QED$_{3}$-GNY model in fixed $d=3$ dimensions; 
as in the previous section, there are no additional contributions specific to $d=3$. The large-$N_{f}$ result is shown in Fig.~\ref{fig:delCDW_d3} alongside the Pad\'e approximants for the $\epsilon$-expansion results; good agreement is found with those approximants (except at four-loop order). Numerical values of Pad\'e and Pad\'e-Borel resummations of (\ref{eq:delCDW_largeN}) at $N_f=4,6,8$ are also found in Tables~\ref{tab:PadeN4_LargeNExp}-\ref{tab:PadeN8_LargeNExp}, Appendix~\ref{app:QED3GNlargeNf}.

Finally, as in Sec.~\ref{sec:Kekule_eps}, we can turn off the gauge coupling and study the resulting chiral $O(2)$ GNY model (ungauged NJL model) in the large-$N_f$ expansion. The critical exponents $\eta_{\phi}$ and $1/\nu$ for this model have already been determined to $O(1/N_{f}^2)$ in Ref.~\cite{Gracey1993b} and Refs.~\cite{Gracey1994,Gracey1994b}, respectively. 
Here, we provide the large-$N_{f}$ analysis for the CDW exponent, to $O(1/N_{f})$ in general $2<d<4$. The pertinent quantities are given by
\begin{eqnarray}
\eta_{1} & = & -\frac{\left(\mu-1\right)\Gamma\left(2\mu-1\right)}{\mu\Gamma\left(1-\mu\right)\Gamma\left(\mu\right)^{3}},\\
\chi_{\phi,1} & = & 0,\\
\eta_{\mathcal{O},3^{\prime}} &= & -\frac{1}{\mu-1}\eta_{1}.
\end{eqnarray}
The CDW exponent is computed via the relation $\Delta_{\text{CDW}}^\text{Kekul\'e}=2\mu-1+\eta_{\mathcal{O},3^{\prime}}.$ 
Again, when this quantity is expanded in $d=4-\epsilon$ up to $O(\epsilon^{4})$, we find agreement with the counterpart expression determined in Sec.~\ref{sec:Kekule_eps}, when the latter is expanded in powers of $1/N_{f}$ to $O(1/N_{f}$).
In fixed $d=3$ the result is 
\begin{equation}\label{CDWKekuleLargeN}
\left.\Delta_{\text{CDW}}^\text{Kekul\'e}\right|_{d=3} = 2-\frac{8}{3\pi^{2}N_{f}}.
\end{equation}
This agrees with Ref.~\cite{boyack2019b}, which computed $\Delta_{\overline{\Psi}\Psi}=\Delta_{\text{CDW}}^\text{Kekul\'e}$ in the $O(2)$ GNY model in fixed $d=3$ dimensions; 
as in the previous section, there are no additional contributions specific to $d=3$. The large-$N_{f}$ result is shown in Fig.~\ref{fig:delCDWXY_d3} alongside the Pad\'e approximants for the $\epsilon$-expansion results; there is excellent agreement with the four-loop approximants for values $N_f\gtrsim 6$. Numerical values of Pad\'e and Pad\'e-Borel resummations of (\ref{CDWKekuleLargeN}) for select values of $N_f$ are presented in Table~\ref{tab:XYPade_LargeNExp}, Appendix~\ref{app:GN}.

\subsection{VBS bilinear vs VBS order parameter}

\begin{figure}[t]
\centering\includegraphics[width=0.85\columnwidth]{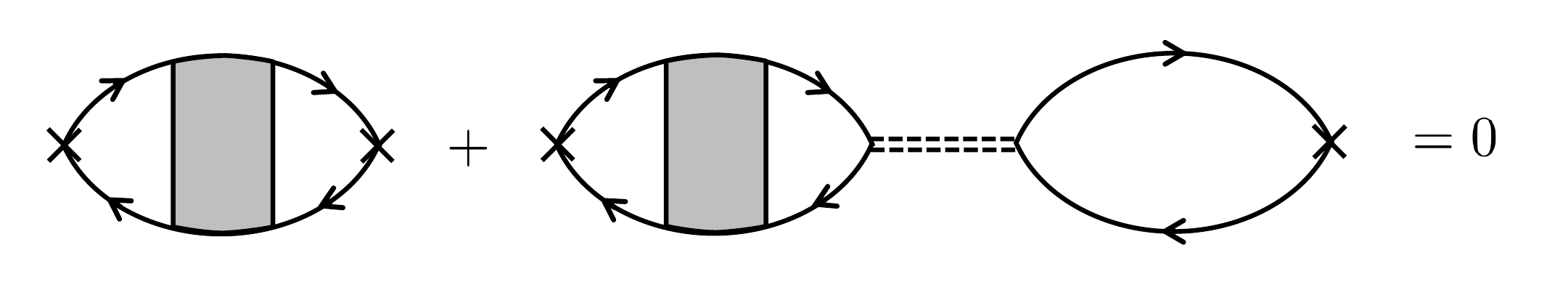}
\caption{Example of cancellation in the two-point correlation function of a VBS bilinear at a given order in the large-$N_f$ expansion.}
\label{fig:dumbbell}
\end{figure}

What about the scaling dimension of the VBS fermion bilinears $i\overline{\Psi}\Gamma_3\Psi$, $i\overline{\Psi}\Gamma_5\Psi$ in the chiral $O(2)$ QED$_3$-GNY (or pure GNY) formulation, or equivalently $\overline{\psi}\psi$, $i\overline{\psi}\gamma_5\psi$ in the gauged (or ungauged) NJL formulation? The order parameter fields $\phi_1,\phi_2$ and those bilinears transform identically under all symmetries, and are in fact not independent operators at the critical fixed point in the large-$N_f$ formalism, as $\phi_1,\phi_2$ arises from the Hubbard-Stratonovich decoupling of a four-Fermi interaction. This is sensible, since VBS correlations at criticality are already controlled by the order parameter anomalous dimension $\eta_\phi$ [see Eqs.~(\ref{VBScorr}) and (\ref{SVBS})]. At the critical fixed point, the above VBS fermion bilinears are in fact set to zero by the equation of motion for $\phi$~\cite{giombi2017,benvenuti2018,benvenuti2019}. Diagrammatically, one finds that the two-point correlation function of a VBS bilinear vanishes order by order in the $1/N_f$ expansion, due to ``dumbbell'' diagrams (Fig.~\ref{fig:dumbbell}, where the double dashed line denotes the large-$N_f$ scalar-field propagator). The cancellation follows from the fact that in the long-wavelength limit, the large-$N_f$ scalar field propagator is simply (minus) the inverse of the fermion bubble. Note that these dumbbell diagrams are only possible when the operator insertion (denoted by ``$\times$'' in Fig.~\ref{fig:dumbbell}) corresponds to a VBS bilinear, which itself appears in the Yukawa vertex. In Ref.~\cite{janssen2020}, the critical scaling of the VBS correlation function was associated with the dimension of the $i\overline{\Psi}\Gamma_3\Psi$, $i\overline{\Psi}\Gamma_5\Psi$ fermion bilinears; as discussed above and also in Refs.~\cite{giombi2017,benvenuti2018,benvenuti2019}, these bilinears technically correspond to vanishing operators due to the presence of the diagrams in Fig.~6. Thus the correct asymptotic scaling of the VBS correlation function is dictated by $\eta_\phi$.

\section{Conclusions}

In summary, we have computed several critical exponents for the VBS transition in lattice QED$_3$ using high-order $\epsilon$- and large-$N_f$ expansions. We have established the emergent $O(2)$ symmetry at the critical point, previously only conjectured, by showing that the only potentially relevant $\mathbb{Z}_4$ anisotropy term is in fact irrelevant in the infrared already at leading order in the $\epsilon$ expansion. We have performed state-of-the-art $O(\epsilon^4)$ computations of several critical exponents in the resulting chiral $O(2)$ QED$_3$-GNY model, equivalent to the gauged NJL model: the anomalous dimension $\eta_\phi$, which controls the power-law decay of VBS two-point correlations at criticality; the correlation length exponent $\nu$; and the exponent $\Delta_\text{CDW}$, which controls the power-law decay of the simplest competing order (CDW order). In the large-$N_f$ expansion, we have newly computed $\eta_\phi$ to $O(1/N_f^2)$. Furthermore, by computing all exponents in $2<d<4$ we have shown that they agree with the $\epsilon$-expansion results at $O(\epsilon^4,1/N_f^p)$, with $p=1,2$ the highest order computed. We have additionally computed the CDW exponent at the Kekul\'e VBS transition on the honeycomb lattice in both $\epsilon$- and large-$N_f$ expansions. Finally, we have performed Pad\'e and Pad\'e-Borel resummations for all critical exponents to obtain numerical estimates for flavor numbers $N_f$ currently accessible to QMC simulations.

\acknowledgements

We acknowledge S. Giombi, Z. Y. Meng, and W. Witczak-Krempa for useful discussions. R.B. was supported by the Theoretical Physics Institute at the University of Alberta. The work of J.A.G. was supported by a DFG Mercator Fellowship and he thanks the Mathematical Physics Group at Humboldt University, Berlin where part of this work was carried out, for hospitality. J.M. was supported by NSERC grant \#RGPIN-2014-4608, the CRC Program, CIFAR, and the University of Alberta.

\appendix
\numberwithin{equation}{section}
\numberwithin{figure}{section}

\onecolumngrid

\section{Resummed critical exponents: VBS transition in lattice QED$_3$}
\label{app:QEDGN}

In the following pages we present tables containing the numerical values of Pad\'e (P) and Pad\'e-Borel (PB) resummations in $d=3$ of the two, three, and four-loop $\epsilon$-expansion results (Sec.~\ref{app:QEDGNepsilon}) and large-$N_f$ expansion results (Sec.~\ref{app:QED3GNlargeNf}) for the critical exponents of the chiral $O(2)$ QED$_3$-GNY model (\ref{eq:XYQEDGN_Model}). Given the power-series expansion $\Delta(\delta) = \sum_{k=0}^{\infty}\Delta_{k}\delta^k$ of a quantity $\Delta$ in terms of a small dimensionless parameter $\delta$ (here $\epsilon$ or $1/N_f$), the Borel sum is defined as 
$B_{\Delta}(\delta) = \sum_{k=0}^{\infty}\Delta_{k}\delta^k/k!$. The Pad\'e-Borel transform is then
\begin{equation}
\Delta(\delta)=\int_{0}^{\infty}dt\,e^{-t}B_{\Delta}(\delta t).
\end{equation}
The $\delta$-expansion coefficients have been computed here only to a finite order, and therefore in the expression above, $B$ is replaced by the appropriate ($\epsilon$-expansion or large-$N_{f}$) Pad\'e approximant.

\clearpage

\subsection{$\epsilon$ expansion}
\label{app:QEDGNepsilon}

We present estimates of $\eta_\phi$, $1/\nu$, and $\Delta_\text{CDW}$ for $N_f=4$ (Table~\ref{tab:PadeN4_EpsilonExp}), $N_f=6$ (Table~\ref{tab:PadeN6_EpsilonExp}), and $N_f=8$ (Table~\ref{tab:PadeN8_EpsilonExp}). Values for which the approximant either has a pole in the domain $\epsilon \in [0,1]$, is undefined, or is negative, are denoted in the tables by $\times$.

\twocolumngrid

\begin{table}[t]
\caption{$\epsilon$-expansion resummations for $N_{f}=4$.}
\label{tab:PadeN4_EpsilonExp}
\centering\begin{tabular}{|c|c|c|c|}
\hline 
 & $\eta_{\phi}$ & $\nu^{-1}$ &  $\Delta_\text{CDW}$ \tabularnewline
\hline 
\hline 
$\text{P}_{[0/2]}$ & $\times$ & 0.691178 &  1.58288 \tabularnewline
\hline 
$\text{PB}_{[0/2]}$ & $\times$ & 0.921858 &  1.83814 \tabularnewline
\hline 
$\text{P}_{[1/1]}$ & 1.75502 & 0.519214  & 1.35423 \tabularnewline
\hline 
$\text{PB}_{[1/1]}$ & 1.75597 & 0.426303  & 1.35013 \tabularnewline
\hline 
$\text{P}_{[0/3]}$ & $\times$ & 0.398022  & 1.5827 \tabularnewline
\hline 
$\text{PB}_{[0/3]}$ & $\times$ & 0.829563  & 1.74961 \tabularnewline
\hline 
$\text{P}_{[1/2]}$ & 3.88237 & $\times$  & 1.5827 \tabularnewline
\hline 
$\text{PB}_{[1/2]}$ & $\times$ & $\times$  & 1.55548 \tabularnewline
\hline 
$\text{P}_{[2/1]}$ & $\times$ & $\times$  & $\times$ \tabularnewline
\hline 
$\text{PB}_{[2/1]}$ & $\times$ & 0.0500637  & $\times$ \tabularnewline
\hline 
$\text{P}_{[0/4]}$ & $\times$ & 1.37123  & 1.34501 \tabularnewline
\hline 
$\text{PB}_{[0/4]}$ & $\times$ & 0.786758  & 1.70129 \tabularnewline
\hline 
$\text{P}_{[1/3]}$ & 0.679069 & 0.541884  & 1.58288 \tabularnewline
\hline 
$\text{PB}_{[1/3]}$ & 1.2536 & 0.553304  & $\times$ \tabularnewline
\hline 
$\text{P}_{[2/2]}$ & 1.97642 & 0.138865  & 1.45687 \tabularnewline
\hline 
$\text{PB}_{[2/2]}$ & $\times$ & $\times$  & 1.45722 \tabularnewline
\hline 
$\text{P}_{[3/1]}$ & 1.94078 & 0.167599  & 1.48159 \tabularnewline
\hline 
$\text{PB}_{[3/1]}$ & 1.97185 & 0.0165785  & 1.49813 \tabularnewline
\hline 
\end{tabular}
\end{table}

\begin{table}[t]
\caption{$\epsilon$-expansion resummations for $N_{f}=6$.}
\label{tab:PadeN6_EpsilonExp}
\centering\begin{tabular}{|c|c|c|c|}
\hline 
 & $\eta_{\phi}$ & $\nu^{-1}$ &  $\Delta_\text{CDW}$ \tabularnewline
\hline 
\hline 
$\text{P}_{[0/2]}$ & $\times$ & 0.795683  & 1.7631 \tabularnewline
\hline 
$\text{PB}_{[0/2]}$ & $\times$ & 1.00854  & 1.95805 \tabularnewline
\hline 
$\text{P}_{[1/1]}$ & 1.47169 & 0.640775  & 1.63251 \tabularnewline
\hline 
$\text{PB}_{[1/1]}$ & 1.47194 & 0.584353  & 1.62754 \tabularnewline
\hline 
$\text{P}_{[0/3]}$ & $\times$ & 0.592311  & 1.74195 \tabularnewline
\hline 
$\text{PB}_{[0/3]}$ & $\times$ & 0.92582  & 1.88102 \tabularnewline
\hline 
$\text{P}_{[1/2]}$ & 1.89107 & $\times$  & 1.73957 \tabularnewline
\hline 
$\text{PB}_{[1/2]}$ & $\times$ & $\times$  & 1.72435 \tabularnewline
\hline 
$\text{P}_{[2/1]}$ & 1.49774 & 0.350477  & $\times$ \tabularnewline
\hline 
$\text{PB}_{[2/1]}$ & 1.49694 & 0.392351  & $\times$ \tabularnewline
\hline 
$\text{P}_{[0/4]}$ & $\times$ & 0.750097  & 1.70513 \tabularnewline
\hline 
$\text{PB}_{[0/4]}$ & $\times$ & 0.8858  & 1.84369 \tabularnewline
\hline 
$\text{P}_{[1/3]}$ & 0.999387 & 0.669572  & $\times$ \tabularnewline
\hline 
$\text{PB}_{[1/3]}$ & 1.25658 & 0.644327  & $\times$ \tabularnewline
\hline 
$\text{P}_{[2/2]}$ & 1.55518 & 0.465496  & 1.70842 \tabularnewline
\hline 
$\text{PB}_{[2/2]}$ & $\times$ & 0.429557  & 1.70409 \tabularnewline
\hline 
$\text{P}_{[3/1]}$ & 1.5527 & 0.493025  & 1.72315 \tabularnewline
\hline 
$\text{PB}_{[3/1]}$ & 1.56486 & 0.435965  & 1.72842 \tabularnewline
\hline 
\end{tabular}
\end{table}

\begin{table}[t]
\caption{$\epsilon$-expansion resummations for $N_{f}=8$.}
\label{tab:PadeN8_EpsilonExp}
\centering\begin{tabular}{|c|c|c|c|}
\hline 
 & $\eta_{\phi}$ & $\nu^{-1}$  & $\Delta_\text{CDW}$ \tabularnewline
\hline 
\hline 
$\text{P}_{[0/2]}$ & $\times$ & 0.860757  & 1.84821 \tabularnewline
\hline 
$\text{PB}_{[0/2]}$ & $\times$ & 1.06164  & 2.01877 \tabularnewline
\hline 
$\text{P}_{[1/1]}$ & 1.34162 & 0.714953  & 1.74425 \tabularnewline
\hline 
$\text{PB}_{[1/1]}$ & 1.34238 & 0.675792  & 1.74029 \tabularnewline
\hline 
$\text{P}_{[0/3]}$ & $\times$ & 0.709312  & 1.82006 \tabularnewline
\hline 
$\text{PB}_{[0/3]}$ & $\times$ & 0.984411  & 1.94698 \tabularnewline
\hline 
$\text{P}_{[1/2]}$ & 1.50719 & $\times$  & 1.81517 \tabularnewline
\hline 
$\text{PB}_{[1/2]}$ & $\times$ & $\times$  & 1.8052 \tabularnewline
\hline 
$\text{P}_{[2/1]}$ & 1.37741 & 0.537753  & 2.23623 \tabularnewline
\hline 
$\text{PB}_{[2/1]}$ & 1.37508 & 0.564138  & $\times$ \tabularnewline
\hline 
$\text{P}_{[0/4]}$ & $\times$ & 0.762125  & 1.80981 \tabularnewline
\hline 
$\text{PB}_{[0/4]}$ & $\times$ & 0.946503  & 1.91307 \tabularnewline
\hline 
$\text{P}_{[1/3]}$ & 1.1047 & 0.746421  & 1.80381 \tabularnewline
\hline 
$\text{PB}_{[1/3]}$ & 1.22268 & 0.714258  & 1.80322 \tabularnewline
\hline 
$\text{P}_{[2/2]}$ & 1.37998 & 0.618923  & 1.80761 \tabularnewline
\hline 
$\text{PB}_{[2/2]}$ & 1.38603 & 0.601702  & 1.80332 \tabularnewline
\hline 
$\text{P}_{[3/1]}$ & 1.37997 & 0.642166  & 1.81526 \tabularnewline
\hline 
$\text{PB}_{[3/1]}$ & 1.38574 & 0.614338  & 1.81724 \tabularnewline
\hline 
\end{tabular}
\end{table}

\clearpage

\onecolumngrid

\subsection{Large-$N_f$ expansion}
\label{app:QED3GNlargeNf}

Here $1/N_f$ is treated as the small expansion parameter for resummation. We present estimates of $\eta_\phi$, $1/\nu$, and $\Delta_\text{CDW}$ for $N_f=4$ (Table~\ref{tab:PadeN4_LargeNExp}), $N_f=6$ (Table~\ref{tab:PadeN6_LargeNExp}), and $N_f=8$ (Table~\ref{tab:PadeN8_LargeNExp}). Approximants which are either singular in the domain $N_{f}\geq1$, undefined, or negative, are denoted by $\times$. 
The exponents that are unknown beyond $O\left(1/N_{f}\right)$ are denoted by --; for these quantities only one approximant can be used.

As both Pad\'e and Pad\'e-Borel resummations fail for $\eta_\phi$, we have performed resummations of $\eta_\phi^{-1}$, then taken the reciprocal, which is denoted by $1/\eta_\phi^{-1}$ in the tables.

\begin{table}[h]
\caption{$1/N_{f}$-expansion resummations for $N_f=4$.}
\label{tab:PadeN4_LargeNExp}
\begin{tabular}{|c|c|c|c|c|}
\hline 
 & $\eta_{\phi}$ & $1/\eta_{\phi}^{-1}$ & $\nu^{-1}$ & $\Delta_\text{CDW}$ 
 \tabularnewline
\hline 
\hline 
$\text{P}_{[0/1]}$ & $\times$ & 1.47283 & 0.596846 & 1.71106 \tabularnewline
\hline 
$\text{PB}_{[0/1]}$ & $\times$ & 1.36619 & 0.670006 & 1.74054 \tabularnewline
\hline 
$\text{P}_{[0/2]}$ & $\times$ & 1.87442 & -- & -- \tabularnewline
\hline 
$\text{PB}_{[0/2]}$ & $\times$ & 1.5627 & -- & -- \tabularnewline
\hline 
$\text{P}_{[1/1]}$ & $\times$ & $\times$ & -- & -- \tabularnewline
\hline 
$\text{PB}_{[1/1]}$ & $\times$ & $\times$ & -- & -- \tabularnewline
\hline 
\end{tabular}
\end{table}

\begin{table}[h]
\caption{$1/N_{f}$-expansion resummations for $N_f=6$.}
\label{tab:PadeN6_LargeNExp}
\begin{tabular}{|c|c|c|c|c|}
\hline 
 & $\eta_{\phi}$ & $1/\eta_{\phi}^{-1}$ & $\nu^{-1}$ & $\Delta_\text{CDW}$ \tabularnewline
\hline 
\hline 
$\text{P}_{[0/1]}$ & $\times$ & 1.31522 & 0.689505 & 1.79763 \tabularnewline
\hline 
$\text{PB}_{[0/1]}$ & $\times$ & 1.25904 & 0.739952 & 1.81347 \tabularnewline
\hline 
$\text{P}_{[0/2]}$ & $\times$ & 1.4937 & -- & -- \tabularnewline
\hline 
$\text{PB}_{[0/2]}$ & $\times$ & 1.37268 & -- & -- \tabularnewline
\hline 
$\text{P}_{[1/1]}$ & $\times$ & $\times$ & -- & -- \tabularnewline
\hline 
$\text{PB}_{[1/1]}$ & $\times$ & $\times$ & -- & -- \tabularnewline
\hline 
\end{tabular}
\end{table}

\begin{table}[h]
\caption{$1/N_{f}$-expansion resummations for $N_f=8$.}
\label{tab:PadeN8_LargeNExp}
\begin{tabular}{|c|c|c|c|c|c|}
\hline 
 & $\eta_{\phi}$ & $1/\eta_{\phi}^{-1}$ & $\nu^{-1}$ & $\Delta_\text{CDW}$ \tabularnewline
\hline 
\hline 
$\text{P}_{[0/1]}$ & $\times$ & 1.23642 & 0.747531 & 1.84428 \tabularnewline
\hline 
$\text{PB}_{[0/1]}$ & $\times$ & 1.20138 & 0.784343 & 1.85417 \tabularnewline
\hline 
$\text{P}_{[0/2]}$ & $\times$ & 1.33681 & -- & -- \tabularnewline
\hline 
$\text{PB}_{[0/2]}$ & $\times$ & 1.27668 & -- & -- \tabularnewline
\hline 
$\text{P}_{[1/1]}$ & $\times$ & $\times$ & -- & -- \tabularnewline
\hline 
$\text{PB}_{[1/1]}$ & $\times$ & $\times$ & -- & -- \tabularnewline
\hline 
\end{tabular}
\end{table}

\clearpage

\section{Resummed critical exponents: Kekul\'e VBS transition on the honeycomb lattice}
\label{app:GN}

We also present Pad\'e and Pad\'e-Borel resummations in $d=3$ of the CDW exponent $\Delta_\text{CDW}$ in the chiral $O(2)$ GNY model, which describes the Kekul\'e VBS transition on the honeycomb lattice. Table~\ref{tab:XYPade_EpsilonExp} contains resummations of the $\epsilon$-expansion results, and Table~\ref{tab:XYPade_LargeNExp} those of the $1/N_f$-expansion result.

\begin{table}[h]
\caption{$\epsilon$-expansion resummations of $\Delta_\text{CDW}$ for the chiral $O(2)$ GNY model.}
\label{tab:XYPade_EpsilonExp}
\begin{tabular}{|c|c|c|c|}
\hline 
 & $N_f=2$ & $N_f=3$ & $N_f=4$ \tabularnewline
\hline 
\hline 
$\text{P}_{[0/2]}$ & 1.93026 & 1.9708 & 1.99453 \tabularnewline
\hline 
$\text{PB}_{[0/2]}$ & 2.0986 & 2.12455 & 2.1401 \tabularnewline
\hline 
$\text{P}_{[1/1]}$ & 1.78953 & 1.85484 & 1.89093 \tabularnewline
\hline 
$\text{PB}_{[1/1]}$ & 1.78864 & 1.85465 & 1.89089 \tabularnewline
\hline 
$\text{P}_{[0/3]}$ & 1.85727 & 1.90925 & 1.93841 \tabularnewline
\hline 
$\text{PB}_{[0/3]}$ & 2.02194 & 2.05289 & 2.07122 \tabularnewline
\hline 
$\text{P}_{[1/2]}$ & 1.81701 & 1.8799 & 1.91344 \tabularnewline
\hline 
$\text{PB}_{[1/2]}$ & 1.81796 & 1.88001 & 1.91324 \tabularnewline
\hline 
$\text{P}_{[2/1]}$ & 1.80722 & 1.86641 & 1.89753 \tabularnewline
\hline 
$\text{PB}_{[2/1]}$ & 1.80608 & 1.86555 & 1.8971 \tabularnewline
\hline 
$\text{P}_{[0/4]}$ & 1.81432 & 1.88521 & 1.92088 \tabularnewline
\hline 
$\text{PB}_{[0/4]}$ & 1.98289 & 2.01739 & 2.03755 \tabularnewline
\hline 
$\text{P}_{[1/3]}$ & 1.74667 & 1.86897 & 1.91257 \tabularnewline
\hline 
$\text{PB}_{[1/3]}$ & $\times$ & $\times$ & 1.91177 \tabularnewline
\hline 
$\text{P}_{[2/2]}$ & 1.80019 & 1.87288 & 1.91261 \tabularnewline
\hline 
$\text{PB}_{[2/2]}$ & 1.80114 & 1.87301 & 1.91189 \tabularnewline
\hline 
$\text{P}_{[3/1]}$ & 1.80045 & 1.87299 & 1.91301 \tabularnewline
\hline 
$\text{PB}_{[3/1]}$ & 1.80139 & 1.87339 & 1.91304 \tabularnewline
\hline 
\end{tabular}
\end{table}

\begin{table}[h]
\caption{$1/N_f$-expansion resummations of $\Delta_\text{CDW}$ for the chiral $O(2)$ GNY model.}
\label{tab:XYPade_LargeNExp}
\begin{tabular}{|c|c|c|c|}
\hline 
  & $N_f=2$ & $N_f=3$ & $N_f=4$ \tabularnewline
\hline 
\hline 
$\text{P}_{[0/1]}$ & 1.87345 & 1.91382 & 1.93466 \tabularnewline
\hline 
$\text{PB}_{[0/1]}$ & 1.88021 & 1.91711 & 1.93661 \tabularnewline
\hline 
\end{tabular}
\end{table}

\twocolumngrid

\bibliography{XYQEDGNY_References}

\begin{thebibliography}{70}
\bibitem [{Sup()}]{SuppMat}%
  \BibitemOpen
  \href@noop {} {}\bibinfo {note} {We provide the full analytical four-loop
  expressions at general $N_f$ for the $\beta$ and $\gamma$ functions in the
  files \texttt{XYQED3GNYbetagamma.m}, \texttt{XYQED3GNYbetagamma.inc}, and for
  the critical exponents in the files \texttt{XYQED3GNYexponents.m},
  \texttt{XYQED3GNYexponents.inc}.}\BibitemShut {Stop}%
\end{thebibliography}

\end{document}